\documentclass[a4paper,11pt]{article}
\pdfoutput=1 
\usepackage{jheppub}

\usepackage[normalem]{ulem}
\usepackage{graphicx}
\usepackage{dcolumn}
\usepackage{bm}
\usepackage{epstopdf}
\usepackage{mathrsfs}
\usepackage{amssymb,amsmath,amsfonts,latexsym}
\usepackage{nicefrac}
\usepackage[colorlinks=true,linktocpage=true,linkcolor=blue,citecolor=blue]{hyperref}
\usepackage[utf8]{inputenc}
\usepackage{lipsum}
\usepackage{nicefrac}
\usepackage{slashed}
\usepackage{mathtools}

\def\eff{\text{eff}}

\def\tform{t_\text{f}}

\def\pT{p_{_T}}

\def\Kc{{\cal K}}

\def\Raa{Q_\mathrm{AA}}

\newcommand{\beq}{\begin{eqnarray}}
\newcommand{\eeq}{\end{eqnarray}}
\newcommand{\be}{\begin{eqnarray*}}
\newcommand{\ee}{\end{eqnarray*}}
\newcommand{\bal}{\begin{align}}
\newcommand{\eal}{\end{align}}

\newcommand{\dd}{{\rm d}}
\newcommand{\rme}{{\rm e}}

\newcommand{\nn}{\nonumber\\ }

\makeatletter
\def\tagform@#1{\maketag@@@{\ignorespaces#1\unskip\@@italiccorr}}
\let\orgtheequation\theequation
\def\theequation{(\orgtheequation)}
\makeatother

\begin{document}

\title{Medium-induced cascade in expanding media} 

\author[a]{Souvik Priyam Adhya}
\author[b]{Carlos A. Salgado}
\author[a]{Martin Spousta}
\author[c]{Konrad Tywoniuk}

\affiliation[a]{Institute of Particle and Nuclear Physics, Faculty of Mathematics and Physics, Charles University, V Holesovickach 2, 180 00 Prague 8, Czech Republic}

\affiliation[b]{Instituto Galego de F\'isica de Altas Enerx\'ias IGFAE, Universidade de Santiago de Compostela, E-15782 Galicia-Spain}

\affiliation[c]{Department of Physics and Technology, University of Bergen, 5007 Bergen, Norway}

\emailAdd{souvik@ipnp.mff.cuni.cz}
\emailAdd{carlos.salgado@usc.es}
\emailAdd{martin.spousta@mff.cuni.cz}
\emailAdd{konrad.tywoniuk@uib.no}

\date{\today}
\abstract{
Detailed insight into the interplay between parton energy loss and the way deconfined medium created in heavy-ion collisions expands is of great importance for improving the understanding of the jet quenching phenomenon. In this paper we study the impact of the expansion of deconfined medium on the single-gluon emission spectrum, its resummation and the jet suppression factor ($\Raa$) within the BDMPS-Z formalism. We calculate these quantities for three types of expansion scenarios, namely static, exponentially decaying and Bjorken expanding media. The distribution of medium-induced gluons is calculated using an evolution equation with splitting kernels derived from the gluon emission spectra. A universal behavior of splitting kernels is derived in the regime of soft gluon emissions when evaluated at a common effective evolution time $\tau_\eff$.
  Novel scaling features of the resulting gluon distribution and jet $\Raa$ are discussed. For realistic spectra valid beyond the soft-gluon emission limit, where the results are obtained by a numerical solution of the evolution equation, these features are partially replaced by a scaling expected from considering an averaged jet quenching parameter along the trajectory of propagation.
  Further we show that differences arising from different types of the medium expansion can be to a large extent scaled out by appropriate choice of the quenching parameter.
  Sizable differences among the values of the quenching parameter for different types of medium expansion point to the importance of the medium expansion for precise modeling 
of the jet quenching phenomenon.
  }

\maketitle
\flushbottom

\section{Introduction}
\label{sec:intro}

Measurements of jets in heavy-ion collisions at RHIC and the LHC revealed many interesting results. The production of inclusive jets was found to be strongly suppressed in central heavy-ion collisions with respect to proton-proton collisions \cite{Aad:2014bxa,Aaboud:2018twu,Khachatryan:2016jfl,Acharya:2019jyg} as a direct consequence of parton energy loss. It was shown that the fragmentation pattern of jets is significantly modified in heavy-ion collisions and that the lost energy is transferred to soft particles, predominantly emitted away from the jet axis \cite{Chatrchyan:2011sx,Chatrchyan:2014ava,Aad:2014wha,Aaboud:2017bzv,Aaboud:2018hpb,Sirunyan:2018ncy}. These are only few of important results (for a review of experimental results, see e.g. Ref.~\cite{Connors:2017ptx}) which however clearly demonstrate that the rich phenomenon of jet quenching calls for an accurate theoretical description.

In this paper we study one particular aspect of the jet quenching, namely the impact of the medium expansion on the rate of stimulated radiation and the related medium-induced branching. The starting point of the calculations presented here is the formalism for propagation and radiation in a dense medium within the BDMPS-Z framework \cite{Baier:1996kr,Baier:1996sk,Zakharov:1996fv,Zakharov:1997uu}. This allows to resum multiple interactions with the medium through a Schr\"odinger equation for the relevant in-medium correlator, see \cite{Mehtar-Tani:2013pia,Blaizot:2015lma}. The solution can be obtained via direct numerical evaluation  \cite{CaronHuot:2010bp,Feal:2018sml,Ke:2018jem,Andres:2020vxs} or as an expansion in terms of the medium opacity \cite{Wiedemann:2000za,Gyulassy:2000fs,Sievert:2019cwq}. Currently, we work within the approximation of multiple-soft scattering, also referred to as the ``dipole'' or ``harmonic oscillator'' approximation \cite{Salgado:2003gb,Armesto:2003jh}, when the resummation for dense media can be performed analytically and that describes well the regime of typical gluon emissions \cite{Arnold:2009mr}.\footnote{Improvements to account for rare emissions can also be systematically included \cite{Mehtar-Tani:2019tvy,Mehtar-Tani:2019ygg,Barata:2020sav}.} 

Multiple scattering in expanding media was analyzed in \cite{Baier:1998yf} and later in \cite{Salgado:2002cd,Salgado:2003gb,Zakharov:2007pj} (see also \cite{Feal:2019xfl} for a numerical solution in this case). These calculations indicate an important impact of the finite expanding medium on the observable quantities such as the nuclear modification factor of hadrons. Interesting features, such as the scaling of gluon energy spectra in expanding media with average transport coefficient, were early identified \cite{Salgado:2002cd,Salgado:2003gb}. This scaling indicates that some of the main features of the medium-induced spectra remain unchanged no matter the underlying density profile of the background medium. 

These approaches have been successfully confronted with experimental data on jet and single-inclusive hadron suppression, see for example \cite{Armesto:2011ht,Burke:2013yra,Mehtar-Tani:2014yea,Arleo:2017ntr,Andres:2019eus,Feal:2019xfl}, with the  aim to reliably extract properties of the dense medium created in heavy-ion collisions. Phenomenological studies aim ultimately at establishing the relation between the jet quenching parameter and the energy density of the quark gluon plasma \cite{Borsanyi:2010cj} which, according to perturbative estimates, should scale like $\hat q/T^3 \sim 2 (\epsilon/T^4)^{3/4}$ \cite{Baier:2002tc}.\footnote{See also \cite{Majumder:2007zh} for other ideas.} Since the energy density is expected to change dramatically during the life-time of the system, jet modifications carry an imprint of this evolution. This prompts us to improve the theoretical description of jet quenching in expanding media.

In this work, rather than attempting a full phenomenological description of experimental data, we focus on shedding light on the universal features of radiative energy loss, and deviations from them.
We extend previous studies to obtain single-inclusive gluon spectra and related in-medium emission rates, and use these to obtain the jet suppression factor for three different types of expanding medium. However, we do not attempt to model the fluctuations related to the production point of the jet or its substructure.
The in-medium distributions are found using numerical solution of the evolution equation for gluon emission spectra introduced in ~\cite{Blaizot:2013hx,Blaizot:2013vha} with the 
important input from a unified treatment of expanding media derived in \cite{Arnold:2008iy}. This allows us to study specific properties and scaling of single-inclusive 
gluon emission spectra and jet suppression factor which can be compared to recent measurements done at the LHC.
While many analyses of in-medium evolution so far have focused on static media \cite{Blaizot:2013hx,Mehtar-Tani:2018zba,Fister:2014zxa}, it is also important to establish whether the qualitative features observed there, such as the rapid transfer of energy to low-energy modes \cite{Baier:2000sb,Kurkela:2018vqr}, can be carried over to expanding cases.

Based on the limit of soft gluon emission, or $x \ll 1$ where $x$ is the energy fraction of the emitted gluon, we are able to closely match all the medium expansion scenarios in terms of a rate that turns out to be constant in a rescaled time variable $\tau_\eff$ which is distinct function of the dimensionless combination $\sqrt{\hat q/p} \, L$, where $p$ is the jet energy and $L$ is the in-medium path length, for each case. This leads to a scaling of the resulting distribution of medium-induced gluons when evaluated at the same \emph{effective} time $\tau_\eff$. However, this scaling behavior is partly washed away when the medium-induced cascade is evaluated with the full BDMPS spectrum, valid beyond the soft gluon limit.  It turns out that the scaling properties in this case correspond more closely to the one expected from considering an average jet quenching coefficient $\langle \hat q \rangle$, which was first numerically discovered in \cite{Salgado:2002cd,Salgado:2003gb}.\footnote{It turns out that the difference between the two forms of scaling, when quantified in terms of an ``effective'' $\hat q$, amounts to an overall factor of 2.}
These features affect in turn the resulting jet suppression factor, indicating that the precise shape of the jet spectrum is sensitive to the details of the expansion and dilution of the hot and dense medium created in heavy-ion collisions. Furthermore, the differences of scaling properties in the small and large-$x$ sectors shed new light on the relation between jet suppression and the amount of energy deposited at the temperature scale in the medium.

The paper is organized as follows. Section~\ref{sec:spectrum} introduces emission spectra and rate of emissions in an expanding medium for three different types of 
media and discusses their properties. Section~\ref{sec:rate-eq} provides calculations of medium evolved gluon emission spectra obtained using the 
evolution equation with input rates from Section~\ref{sec:spectrum}. In Section~\ref{sec:moments}, the moments of gluon spectra are calculated allowing to obtain the 
jet suppression factor, $\Raa$, for different types of expanding media. The scaling properties of the jet $\Raa$ with respect to the transport properties of the 
expanding media are discussed. Section~\ref{sec:conclusions} provides a summary and outlook.

\section{Emission spectrum and rate in an expanding medium} 
\label{sec:spectrum}

Calculations of medium-induced gluon radiation in the evolving media presented in this paper 
are done in the limit of multiple soft scatterings and follow the BDMPS-Z formalism 
\cite{Baier:1996kr,Baier:1996sk,Zakharov:1996fv,Zakharov:1997uu}. 
For the purposes of this paper, we only consider gluon branching.
The starting point is the gluon emission spectrum radiated from an initial massless parton with energy $p$ (we only consider gluon splitting at the moment). The final expression can be cast in a general form as \cite{Arnold:2008iy}
\beq
\label{eq:simple-formula}
\frac{\dd I}{\dd z}  = \frac{\alpha_s}{\pi} P(z) \ln|c(0)| \,,
\eeq
where $P(z) \equiv P_{gg}(z) =2 N_c [1-z(1-z)]^2/z(1-z)$ is the Altarelli-Parisi splitting function. The strong coupling constant $\alpha_s$ runs with the typical transverse momentum accumulated during the emission, $k_\perp \sim (z(1-z)p\hat q)^{1/4}$, but in the remainder of the paper we will treat it as a constant, $\alpha_s = 0.14$.

In \autoref{eq:simple-formula}, $c(t)$ is a function that encodes information about the medium and its expansion \cite{Arnold:2008iy}. It is the solution of a differential equation 
\beq
\label{eq:diff-eq-c}
\frac{\dd^2 c(t)}{{\dd t}^2} + \Omega^2(t) c(t) = 0 \,,
\eeq
where $\Omega(t)$ is a time-dependent, complex frequency. For our purposes (gluon splitting), this frequency is simply given by
\beq
\Omega^2(t) = -i\frac{\hat q_{\rm eff}(t)}{2z(1-z)p} \,,
\eeq
where the effective jet quenching parameter is given by $\hat q_{\rm eff}(t) = [1-z(1-z)] \hat q (t)$. The boundary conditions are such that $c(t)$ approaches 1 at $t \to \infty$; this realizes the fact that the particle ends up in a vacuum state, i.e. $\hat q \to 0$ and therefore $\Omega(t) \to 0$ as $t \to \infty$. On the other hand, $t=0$ corresponds to the position of the hard scattering that produces the hard particle sourcing the splitting. The solutions to the differential equation \ref{eq:diff-eq-c} are subject to boundary conditions that, in the case of the spectrum in \autoref{eq:simple-formula} are given by $c(\infty) = 1$ and $\partial_t c(t)|_{t=\infty} = 0$.

We can also derive an emission rate, defined as
\beq 
\mathcal{K}(z,\tau) \equiv \frac{\dd I}{\dd z \dd \tau} \,, 
\eeq 
per unit ``time'' $\tau$. This is a dimensionless number defined as
\beq 
\tau = \sqrt{\frac{\hat q_0}{p}}L \,, 
\label{eq:tau} 
\eeq 
where $L$ is the distance the initial parton travels trough the medium. 
The parameter ${\hat q_0} = {\hat q}(t_0)$ is the initial value of the jet quenching parameter.
The rate $\mathcal{K}(z,\tau)$ is an input to calculations of medium evolved gluon spectra, which will be discussed in \autoref{sec:rate-eq}.

\begin{figure}
\centering
\includegraphics[width=0.5\textwidth]{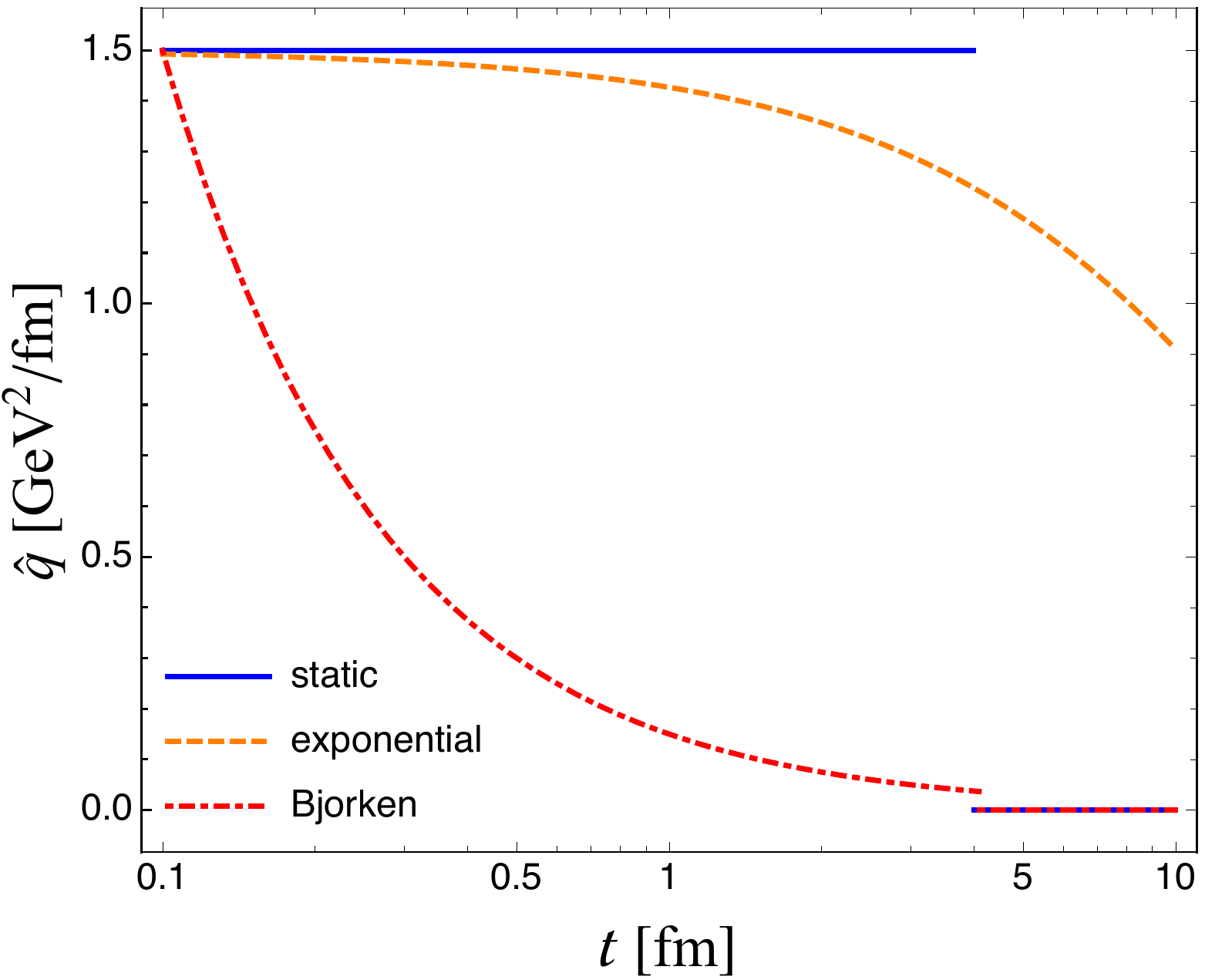}
\caption{The time dependence of the ${\hat q}$ coefficient for the different expanding scenarios considered in this paper: a static medium (blue, solid), exponentially decaying medium (orange, dashed) and the Bjorken expanding medium (red, dot-dashed) with $t_0 =0.1$fm. 
}
\label{fig:qhat}
\end{figure}
For expanding media, the quenching parameter is time dependent, ${\hat q} = {\hat q}(t)$. 
It was early realized that the single-gluon emission spectrum for different medium expansion scenarios possessed scaling features when plotted for the same value of the (properly defined) average transport coefficient \cite{Salgado:2002cd,Salgado:2003gb}, where
the average quenching parameter for a given type of the expanding medium is
\beq
\label{eq:average-qhat}
\langle \hat q \rangle = \frac{2}{L^2} \int_{t_0}^{L+t_0} \dd t \, (t-t_0) \hat q(t) \equiv  \frac{2}{L^2} \langle(t-t_0) \rangle \,,
\eeq
where $t_0$ corresponds to the time-scale for the onset of quenching effects, i.e. $\hat q(t < t_0)=0 $.

In this work we will consider three examples of medium evolution, differing by $\hat q(t)$ profiles and therefore with different $c(0)$. These are the static medium, exponentially decaying medium and the Bjorken expanding medium. The time dependence of ${\hat q}$ for these different scenarios is fully specified later in this section and summarized in \autoref{fig:qhat}. For the former two examples we can safely put $t_0=0$ while for the Bjorken scenario, where the energy density and therefore also $\hat q(t)$ 
diverges at small times, we have to use a finite $t_0$. Moreover, the exponentially decreasing spectrum is extending up to $t = \infty$, which automatically 
regularizes $\hat q(t)$ at late times, and it is therefore natural to define the average jet quenching parameter as
\beq
\label{eq:average-qhat-2}
\langle \hat q \rangle_{\rm exp} = \frac{2}{L^2} \int_0^\infty \dd t \, t \,\hat q(t) \,,
\eeq
in this case. 

In this section, we establish the \emph{exact} scaling features of the single-gluon emission spectrum and rate in a theoretically well-defined limit, namely the soft gluon emission regime. Surprisingly, they turn out to be slightly different than expected from Eqs.~\ref{eq:average-qhat} and \ref{eq:average-qhat-2}. Below we will analyze how this affects the resulting distribution of medium-induced gluons after passing through the medium.
The reference values for the jet quenching parameter at initial time and the size of the medium used in this section are $\hat q_0 = 1.5$ GeV$^2$/fm and $L = 4$ fm, respectively. Emission spectra and rates for these examples as well as their properties are detailed in the remainder of this section.

\subsection{Static medium}
\label{subsec:static}

 For a static medium, $\hat q(t) = \hat q_0$ for $t < L$ and vanishes at later times, and obviously $\langle \hat q\rangle = \hat q_0$ as well. In this case, $\Omega^2(t) = \Omega^2_0$, see below, at $t <L$ and $\Omega^2(t) = 0$ at $t>L$. The spectrum is given by \cite{Baier:1996kr,Baier:1996sk,Zakharov:1996fv,Zakharov:1997uu}
\beq
\label{eq:spectrum-static}
\frac{\dd I}{\dd z} = \frac{\alpha_s}{\pi} P(z)\, \text{Re} \ln \cos \Omega_0 L \,,
\eeq
where 
\beq
\label{eq:omega-static}
\Omega_0 L = \sqrt{\frac{-i}{2}\frac{\hat q_0}{p}}\kappa(z) L = \frac{1-i}{2} \kappa(z) \tau \,,
\eeq 
and $\kappa(z) = \sqrt{[1-z(1-z)]\big/[z (1-z)]}$. Focusing on the small-$z$ limit, $z \ll 1$, and defining the gluon frequency $\omega = z p$, we see that the spectrum has two regimes, namely
\beq
\label{eq:static-asymptotes}
\omega \frac{\dd I}{\dd \omega} \simeq 2\bar \alpha \begin{cases} \sqrt{\frac{\hat q L^2}{4 \omega}} & \text{for}\quad \omega \ll \hat q L^2  \,,\\ \frac{1}{12}\left(\frac{\hat q L^2}{2\omega} \right)^2 & \text{for} \quad \omega \gg \hat q L^2 \,,  \end{cases}
\eeq
where $\bar \alpha \equiv \alpha_s N_c/\pi$. The characteristic (hard) gluon frequency is often denoted as $\omega_c = \frac{1}{2}\hat q L^2$.
The $\omega^{-1/2}$ behavior at low energies is a consequence of the LPM interference effect, and applies for gluons with formation times shorter than the medium length, $\tform \sim \sqrt{\omega/\hat q } < L$. This essential feature fundamentally impacts the resulting distribution of medium-induced parton cascade \cite{Baier:2000mf,Blaizot:2013hx}. Furthermore, in this regime the spectrum is proportional to the in-medium path length, $\omega \dd I/\dd \omega \propto L$. At long formation times, $\tform \sim \sqrt{\omega / \hat q } > L$, or $\omega > \omega_c$, the spectrum is strongly suppressed.

Returning now to finite $z$, in terms of the evolution variable $\tau$, the rate then becomes
\beq
\label{eq:rate-static}
\mathcal{K}(z,\tau) = \frac{\alpha_s}{2\pi} P(z) \kappa(z)\, \text{Re} \left[(i-1) \tan \left(\frac{1-i}{2} \kappa(z) \tau \right) \right]\,.
\eeq
It is also useful to recall the ``soft'' limit of this spectrum that will be used for comparison later on. We are interested in the regime $\kappa(z) \tau\sim \tau/\sqrt{z} \gg 1$, for large $\tau$ or for $z \ll \tau^2$, where we can expand the cosine in \autoref{eq:spectrum-static} to obtain
\beq
\label{eq:spectrum-static-soft}
\left.\frac{\dd I}{\dd z} \right|_{\tau \gg \sqrt{z}} \simeq \frac{\alpha_s}{2\pi} P(z) \kappa(z) \tau \,,
\eeq
where the rate ${\cal K}(z,\tau) \simeq \frac{\alpha_s}{2\pi}P(z) \kappa(z)$ is constant in ``time''. In order to highlight the features of the rate, and corresponding distribution of gluons emitted in the medium, we will further simplify this expression by neglecting all $z$-dependence apart from the (apparent) singular behavior in $z\to 0$ and $z \to 1$. 
In this case, the rate reads
\beq
\label{eq:rate-static-soft}
\left.{\cal K}(z,\tau) \right|_{\rm sing} = \frac{\bar \alpha}{[z(1-z)]^{3/2}} \,.
\eeq
This can also be found by considering the limit of large times $\tau$, conversely small $z$, directly in \autoref{eq:rate-static}, where $\lim_{x \to \infty}\tan (1-i)x = -i$ and hence the rate tends to constant, time-independent value at large times. It turns out that the medium evolution of the gluon distribution is exactly solvable using \autoref{eq:rate-static-soft} \cite{Blaizot:2013hx}, which makes it an interesting limiting case.

\subsection{Exponentially decaying medium} 
\label{subsec:expo}
 For exponentially decaying media the profile of the jet quenching parameter is given by
\beq
\label{eq:definition-exp}
\hat q(t) = \hat q_0 {\rm e}^{-t/L}.
\eeq
Note that in this case the average parameter, according to \autoref{eq:average-qhat-2}, is $\langle \hat q\rangle_{\rm exp} = 2 \hat q_0$, i.e. twice as big as for the static medium. This is a consequence of the fact that, although exponentially suppressed, the quenching is allowed to take place over very long distances.

The solution of $c(t)$ satisfying the boundary conditions at $t \to \infty$ is readily found, and in this case the spectrum is given by
\beq
\frac{\dd I}{\dd z} = \frac{\alpha_s}{\pi} P(z)\, \text{Re} \ln  J_0( 2\Omega_0 L ) \,, 
\label{eq:spectrum-expo}
\eeq
where $J_0(z)$ is a Bessel function of the first kind and $\Omega_0 L$ is given in \autoref{eq:omega-static}. We point out the factor 2 appearing inside the Bessel function, that highlights some of the peculiar features of this particular scenario.
The rate then becomes
\beq
\label{eq:rate-expo}
\mathcal{K}(z,\tau) = \frac{\alpha_s}{\pi} P(z) \kappa(z) \,\text{Re} \left[ (i-1) \frac{J_1\big((1-i)\kappa(z) \tau \big)}{J_0\big((1-i)\kappa(z) \tau \big)} \right] \,.
\eeq
We notice again the ratio of Bessel functions tend to a constant value at large times $\lim_{x\to \infty} J_1\big((1-i)x \big) \big/J_0\big((1-i)x \big) = -i$. However, given the profile defined in \autoref{eq:definition-exp}, this limiting value is twice as large as for the static case,
\beq
\lim_{\tau \to \infty} {\cal K}_{\rm exp}(z,\tau) = 2 \lim_{\tau \to \infty} {\cal K}_{\rm static}(z,\tau) \,.
\eeq
This mismatch can in principle be remedied by rescaling medium parameters in \autoref{eq:definition-exp}, e.g. $L \to L/2$.

However, to put these insights onto firmer theoretical ground and reveal the scaling features of the spectrum and rate, let us presently analyze the leading behavior arising in the limit of 
$z \rightarrow 0$ and $z \rightarrow 1$. We will therefore approximate $\kappa(z)\approx 1/\sqrt{z(1-z)}$ and $P(z) \approx 2N_c/(z(1-z))$, and employ the asymptotic form of the Bessel functions for large arguments.
With these approximations the emission spectrum \ref{eq:spectrum-expo} can be written as,
\begin{align}
\label{eq:spectrum-expo-approx}
\left.\frac{\dd I}{\dd z} \right|_{\rm sing} &\simeq \frac{2\bar \alpha}{z(1-z)}\, \left\{\text{Re} \ln \cos\left[(1-i)\sqrt{\frac{ \hat q_0 L^2}{ z(1-z)p}} -\frac{\pi}{4} \right] +\frac{1}{4}\ln \left(\frac{2 z(1-z)p}{\pi^2 \hat q_0 L^2} \right) \right\} \nn
&\approx \frac{2\bar \alpha}{z(1-z)} \, \text{Re} \ln \cos\left[\frac{1-i}{2}\sqrt{\frac{4 \hat q_0 L^2}{z(1-z)p}} -\frac{\pi}{4} \right] \,,
\end{align}
where in the second step we neglected the second term in the small or large $z$ limit, as indicated by the subscript ``sing'' next to the spectrum.
 We observe that, apart from the factor $\pi/4$ under the cosine, there is only a factor $4$ difference under the square root between the exponentially decaying medium and a static one, cf. \autoref{eq:spectrum-static}.
This is also a factor 2 bigger than the expected scaling by using the average $\hat q$ parameter introduced in \autoref{eq:average-qhat-2}.

Based on \autoref{eq:spectrum-expo}, we immediately find the asymptotes of the spectrum in the soft and hard limits. For now we will treat $z \ll 1$ with $\omega \equiv z p$. In analogy to \autoref{eq:static-asymptotes}, we find 
\beq
\label{eq:exp-asymptotes}
\omega \frac{\dd I}{\dd \omega} = 2\bar \alpha \begin{cases} \sqrt{\frac{\hat q_0 L^2}{\omega}} & \text{for } \omega \ll \hat q_0 L^2 \\ \frac{1}{16} \left(\frac{\hat q_0 L^2}{\omega} \right)^2 & \text{for } \omega \gg \hat q_0 L^2\end{cases} \,.
\eeq
We therefore affirm that there is no unique rescaling of the medium parameters in \autoref{eq:exp-asymptotes} such as to exactly recover the static spectrum for the whole range of $\omega$ values, cf. \autoref{eq:static-asymptotes}. In particular, we notice that rewriting the exponential spectrum by means of $\langle \hat q\rangle_\text{exp}$ results in a constant mismatch with the static spectrum, i.e.
\beq
\label{eq:scaling-exp-average}
\frac{\dd I_\text{exp}(\langle \hat q \rangle)}{\dd \omega} \Bigg/ \frac{\dd I_\text{static}(\langle \hat q \rangle)}{\dd \omega} = \begin{cases} \sqrt{2} & \text{for } \omega \ll \langle \hat q \rangle L^2 \\ \frac{3}{4} & \text{for } \omega \gg \langle \hat q \rangle L^2 \end{cases} \,.
\eeq
On the other hand, if we instead define an effective, rescaled $\hat q_\eff = 4 \hat q_0$, the 
\beq
\label{eq:scaling-exp-soft}
\frac{\dd I_\text{exp}(\hat q_\eff)}{\dd \omega} \Bigg/ \frac{\dd I_\text{static}(\hat q_\eff)}{\dd \omega} = \begin{cases} 1 & \text{for } \omega \ll \hat q_\eff  L^2 \\ \frac{3}{16} & \text{for } \omega \gg \hat q_\eff L^2 \end{cases} \,,
\eeq
where we note the perfect scaling in the soft sector.
On the other hand, these results signal the breakdown of na\"ive scaling laws that apply to the full kinematical range.

Turning now to the rate in \ref{eq:rate-expo}, we notice that it can be approximated as
\begin{align}
\label{eq:rate-expo-approx}
\left. \mathcal{K}(z,\tau) \right|_{\rm sing} &\simeq \frac{2\bar \alpha}{[z(1-z)]^{3/2}} \, \text{Re} \,(i-1) \tan \left[(1-i)\frac{\tau}{\sqrt{z(1-z)}} -\frac{\pi}{4} \right]\,, \nn
&\approx \frac{2\bar\alpha}{[z(1-z)]^{3/2}}\,,
\end{align}
where in the  second step we additionally assumed that $z \ll \tau^2$.
When comparing with \autoref{eq:rate-static-soft}, we note an overall factor 2 difference of the rate. 
This factor can be absorbed into a redefinition of the evolution time, by defining
\beq
\tau_\eff = 2 \tau \,.
\eeq
In terms of the re-scaled time-variable $\tau_\eff$, the rate is constant,
\beq
\left. \mathcal{K}(z,\tau_\eff) \right|_{\rm sing} = \frac{\bar \alpha}{[z(1-z)]^{3/2}} \,,
\eeq
and, moreover, coincides with the static rate (written per unit time $\tau_\eff = \tau$). Although this seems to be a somewhat artificial manipulation at this stage, we will see its usefulness in making sense out of the Bjorken scenario, discussed next.

\subsection{Bjorken expanding medium} 
\label{subsec:bjorken}

This type of the medium is motivated by the Bjorken expansion, which leads to the drop of energy density $\varepsilon(t)$ with proper time as $\varepsilon(t) = \varepsilon(t_0) (t_0/t)^{4/3}$ for massless relativistic particles. Since $\hat q \propto \varepsilon^{3/4}$, one can therefore model the time dependence of the jet quenching parameter as \cite{Baier:1998yf},
\beq
\hat q(t) = \begin{cases} 0 & {\rm for } \quad t<t_0 \,, \\ \hat q_0 (t_0/t)^\alpha & {\rm for} \quad t_0 < t < L+t_0 \,, \\ 0 & {\rm for} \quad L+t_0 < t \,.\end{cases}
\eeq
The $\alpha \neq 1$ generalizes the above-mentioned Bjorken expansion. We find that $\langle \hat{q}\rangle_{\rm Bjork} = 2\hat{q}_0 t_0/L $ and the average value of $\langle \hat q \rangle$ is in this case dependent on the ratio $t_0/L$. For the typical values $t_0 =0.1$ fm, $L = 4 $ fm, and $\alpha = 1$ we find $\langle \hat q \rangle_{\rm Bjork}/\hat q_0 \approx 0.05$, i.e. the expansion reduces the average quenching parameter by a factor of 20.

The spectra for generic power-law expansions characterized by $\alpha$ were analyzed in \cite{Baier:1998yf,Salgado:2003gb,Arnold:2008iy}. The spectrum is given by
\beq
\label{eq:spectrum-BJ}
\frac{\dd I}{\dd z} = \frac{\alpha_s}{\pi} P(z)\, \text{Re} \ln\left[ \left(\frac{t_0}{L+t_0} \right)^{1/2} \frac{J_{\nu}(z_0)Y_{\nu-1}(z_L) - Y_{\nu}(z_0) J_{\nu-1}(z_L)}{J_\nu(z_L)Y_{\nu-1}(z_L) - Y_{\nu}(z_L) J_{\nu - 1}(z_L)} \right]\,, 
\eeq
for $\alpha < 2$ with $\nu \equiv 1/(2-\alpha)$, and where
\begin{align}
z_0 &\equiv 2\nu \,\frac{1-i}{2} \kappa(z)\sqrt{\frac{\hat q_0}{p}} t_0 = \nu\, (1-i) \kappa(z) \tau_0 \,, \\
z_L &\equiv 2\nu\, \frac{1-i}{2} \kappa(z)\sqrt{\frac{\hat q_0}{p}} \sqrt{t_0 \,(L+t_0)} = \nu\, (1-i) \kappa(z) \sqrt{\tau_0 (\tau + \tau_0)} \,,
\end{align}
where $\tau_0 = \sqrt{\hat q_0/p} t_0$. 

In what follows $\alpha = 1$ (and $\nu = 1$) implying 
\beq
\frac{\dd I}{\dd z} = \frac{\alpha_s}{\pi} P(z)\, \text{Re} \ln\left[ \left(\frac{t_0}{L+t_0} \right)^{1/2} \frac{J_1(z_0)Y_0(z_L) - Y_1(z_0) J_0(z_L)}{J_1(z_L)Y_0(z_L) - Y_1(z_L) J_0(z_L)} \right]\,, 
\eeq
and the rate becomes
\begin{align}
\label{eq:rate:bjorken}
\mathcal{K}(z,\tau) = \frac{\alpha_s}{2\pi} P(z) \kappa(z)\sqrt{\frac{\tau_0}{\tau + \tau_0}} \text{Re} \left[ (1-i) \frac{J_1(z_L) Y_1(z_0) - J_1(z_0) Y_1(z_L) }{J_1(z_0) Y_0(z_L) - J_0(z_L) Y_1(z_0)} \right].
\end{align}
We point out that this rate depends explicitly on $\tau_0$ as well as on $\tau$. The long-time behavior of this scenario stands out compared to the other two cases analyzed above. While the factor inside the square brackets in \autoref{eq:rate:bjorken} goes to a constant, i.e. $\lim_{\tau \to \infty}{\rm Re}[\ldots] = 1$, the square root in front leads to a power-like decay of the rate at large times, i.e.
\beq
\lim_{\tau \to \infty} {\cal K}(z,\tau) = \frac{\alpha_s}{2\pi} P(z) \kappa(z) \sqrt{\frac{\tau_0}{\tau}} \,.
\eeq
However, this can be also obtained for sufficiently small $z$, i.e. $z \ll z_c \equiv \tau_0 \tau$. In fact, for these small $z$ values the properties of the Bjorken expanding and the static case, where $z_c \equiv \tau^2$ are quite similar, i.e. ${\cal K}_{\rm static}(z_c,\tau) \approx {\cal K}_{\rm Bjork}(z_c,\tau) \propto \tau^{-1}$.

Once again, we now turn to the ``singular'' behavior, see the previous sub-section, of the spectrum and rate in order to extract the scaling features. We employ the asymptotic forms of the Bessel functions to calculate the emission spectra for the Bjorken medium \autoref{eq:spectrum-BJ} in the limit of small or large $z$. We finally get
\begin{align}
\label{eq:spectrum-Bjork-approx}
\left.\frac{\dd I}{\dd z} \right|_{\rm sing} & \simeq \frac{2\bar \alpha}{z(1-z)}\, \text{Re} \ln\left[\sqrt{\frac{t_0}{L+t_0}} \cos \left((1-i)\sqrt{\frac{\hat q_0 t_0 L}{z(1-z)p}}\right) \right] \,, \nn
&\simeq \frac{2\bar \alpha}{z(1-z)}\, \left[ \text{Re} \ln \cos\left( \frac{1-i}{2}\sqrt{\frac{4 \hat q_0 t_0  L}{z(1-z)p}} \right)+ \frac{1}{2} \ln \frac{t_0}{L} \right]  \,.
\end{align}
In this case, the spectrum contains an additive term when compared to its equivalent static expression \ref{eq:spectrum-static}. In addition, we see a similar factor of 2 mismatch of the average $\hat q$ as for the exponential case.

Similarly to the derivations above, the asymptotes of the spectrum in the $z \ll 1$ limit are found to be
\beq
\label{eq:bjork-asymptotes}
\omega \frac{\dd I}{\dd \omega} = 2\bar \alpha \begin{cases} \sqrt{\frac{\hat q_0 t_0 L}{\omega}} & \text{for } \omega \ll \hat q_0 L^2 \\ \frac{1}{16} \left(\frac{\hat q_0 t_0 L}{\omega} \right)^2 & \text{for } \omega \gg \hat q_0 L^2\end{cases} \,.
\eeq
Curiously, the remaining mismatch with the static spectrum after rewriting the spectrum in terms of $\langle \hat q \rangle_\text{Bjork}$ are exactly the same as in the exponential case, see Eqs.~\ref{eq:scaling-exp-average} and \ref{eq:scaling-exp-soft}. We note that, strictly speaking, there is no scaling law that can accommodate the behavior of the spectrum for the full range of gluon energies.

Similarly, the rate \ref{eq:rate:bjorken} can be approximated as,
\begin{align}
\label{eq:rate-Bjork-approx}
\left.\mathcal{K}(z,\tau)\right|_{\rm sing} &\simeq \frac{\bar \alpha}{[z(1-z)]^{3/2}} \,\sqrt{\frac{\tau_0}{\tau_0+\tau}} \text{Re} (i-1) \tan\left[ (1-i)\sqrt{\frac{\tau_0 \tau}{z(1-z)}} \right]\,, \nn 
&\approx \frac{\bar \alpha}{ [z(1-z)]^{3/2}} \,\sqrt{\frac{\tau_0}{\tau}} \,,
\end{align}
where in the second line we additionally assumed $z \ll \tau_0 \tau$.
Comparing the above equation with \ref{eq:rate-static}, we note several differences. Overall, similarly to the exponential case, there is a factor of 2 difference in the argument of the tangent with respect to the static case. Additionally, the rate depends on $\tau$ in a different way than in the static case. This is manifested in the extra pre-factor $\sim \sqrt{\tau_0/\tau}$ and the factor $\sqrt{\tau_0 \tau}$ in the argument of the tangent. This additional factor will indeed break the na\"ive scaling of the Bjorken rate with the static and exponential media. 

However, the additional time-dependent pre-factor can be absorbed into a redefinition of the evolution time. Introducing an effective evolution time
\beq
\tau_\eff = 2\sqrt{\tau_0 \tau} \,,
\eeq
with $\dd \tau_\eff = \sqrt{\tau_0/\tau} \dd \tau$, 
we can recast the rate as
\beq
\label{eq:rate-Bjork-approx-2}
\left.\Kc(z,\tau_\eff) \right|_{\rm sing} = \frac{\bar \alpha}{[z(1-z)]^{3/2}} \,.
\eeq
 Generally, in this class of expanding scenarios, the effective time variable can be cast as
\beq
\tau_\text{eff} = \int_0^t \dd t' \sqrt{\frac{\hat q(t')}{p }} \,,
\eeq
for $\alpha < 2$.\footnote{We thank Y. Mehtar-Tani for pointing out this relation.}
In summary, although the rate for the Bjorken expanding medium at first glance leads to a qualitatively different time-dependence (which primarily is not constant in time $\sim \sqrt{\hat q_0/p}\,t$), we can absorb these differences into a clever choice of evolution variable in the singular case. In terms of the new time-variable $\tau_\eff$ the rate is constant and equivalent to the static one.

\subsection{Properties of the emission spectrum and rate}
\label{sec:rate_properties}

\begin{figure}
\centering
\includegraphics[width=0.48\textwidth]{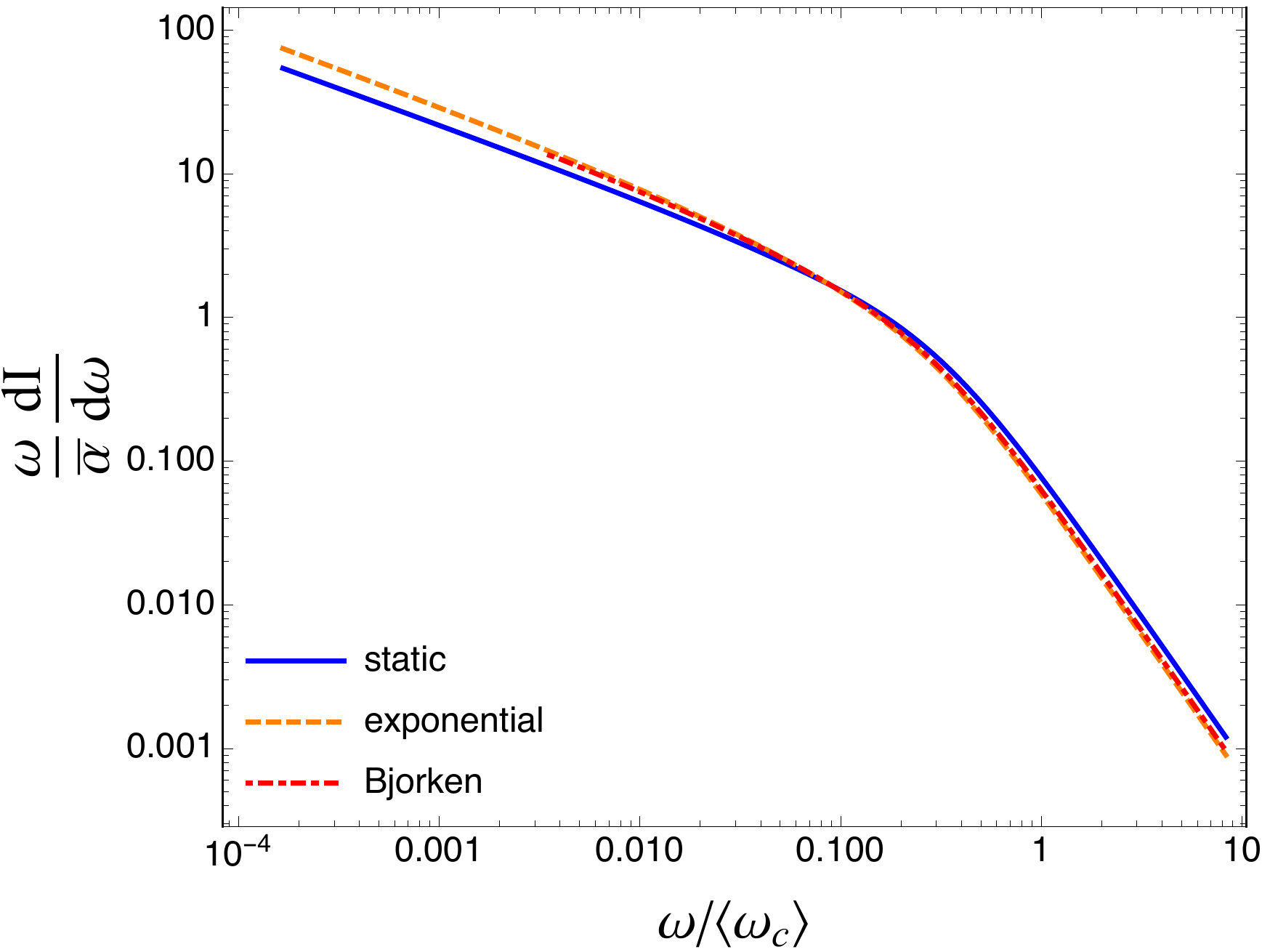}
\includegraphics[width=0.48\textwidth]{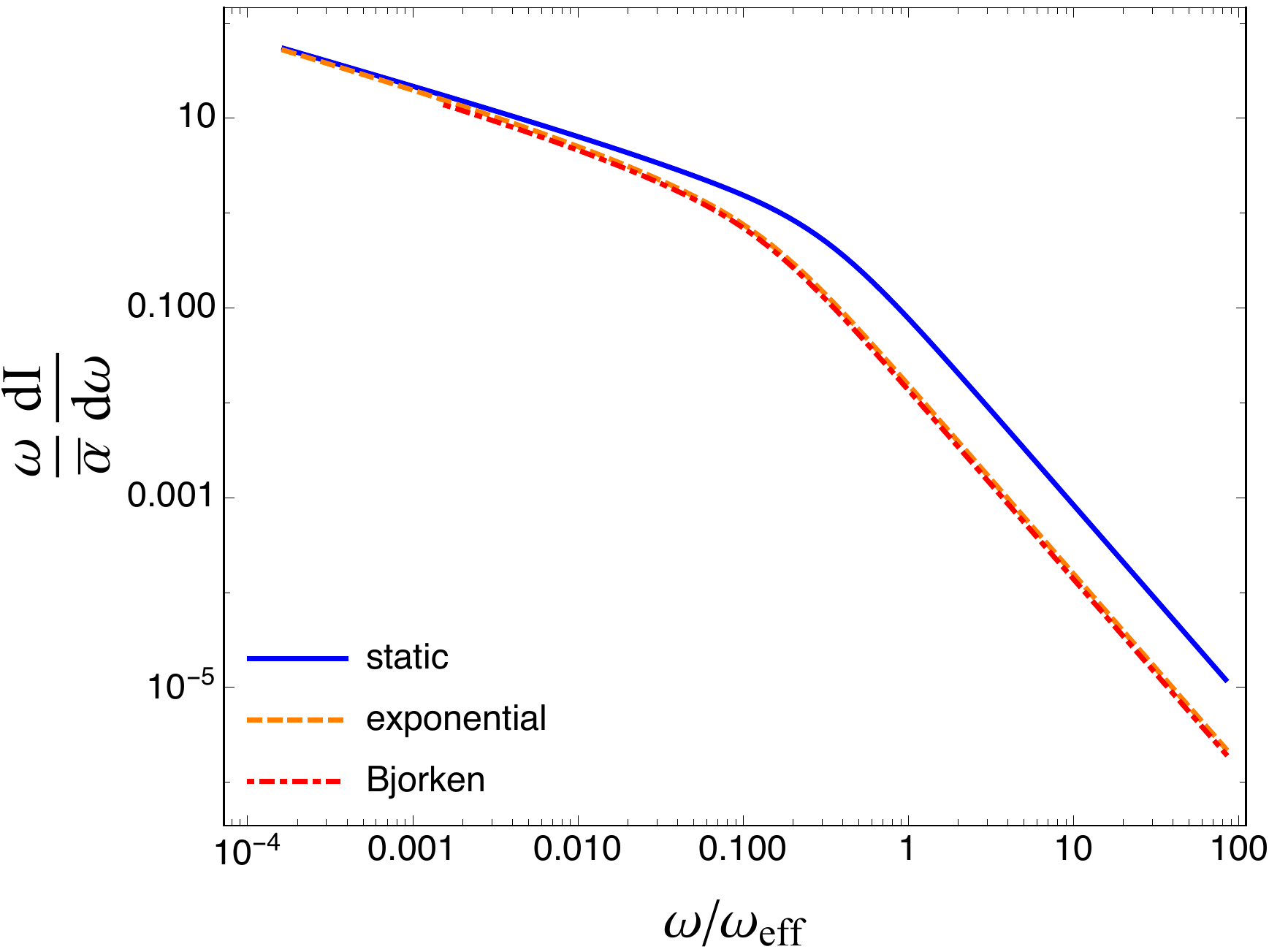}
\caption{Spectrum of medium induced gluons $\omega\, \dd I/\dd \omega$ (in the limit $z \ll 1$) scaled by $\bar\alpha = \alpha_s N_c/\pi$ and plotted as a function of the gluon energy rescaled by $\langle \omega_c \rangle = \langle \hat q \rangle L^2/2$ (left panel) and by $\omega_\text{eff} = \hat q_\eff L^2/2$ (right panel), see definition in the text. For the Bjorken case, we have chosen $t_0=0.1$ fm. The plotting options are the same as in \autoref{fig:qhat}.}
\label{fig:spectra-scaling}
\end{figure}
We compare the spectra of medium-induced gluon in \autoref{fig:spectra-scaling}. In the left panel, we have plotted the spectrum $\omega \,\dd I/\dd \omega$ versus $\omega/\langle \omega_c \rangle$, i.e. the energy rescaled by the maximal available gluon energy in the medium $\langle \omega_c \rangle \equiv \langle \hat q\rangle L^2/2$. We see that the Bjorken model (red, dot-dashed curve in \autoref{fig:spectra-scaling}) approximately respects the scaling, as first discussed in \cite{Salgado:2002cd,Salgado:2003gb}. The exponential profile, with $\langle \hat q \rangle$ defined as in \autoref{eq:average-qhat-2}, also obeys the scaling approximately, cf. the (orange) dashed curve in \autoref{fig:spectra-scaling}. 
The scaling in the soft sector is clearly improved in the right panel of \autoref{fig:spectra-scaling}, where we re-scale by the analytically motivated parameter $\omega_\text{eff}$, which was derived above to be
\beq
\label{eq:omega-eff}
\omega_\text{eff} = \begin{cases} \frac{1}{2} \hat q_0 L^2 & \text{static medium} \\
2 \hat q_0 L^2 & \text{exponentially expansion} \\
2 \hat q_0 t_0 L & \text{Bjorken expansion} \end{cases} \,.
\eeq

\begin{figure}
\centering
\includegraphics[width=0.48\textwidth]{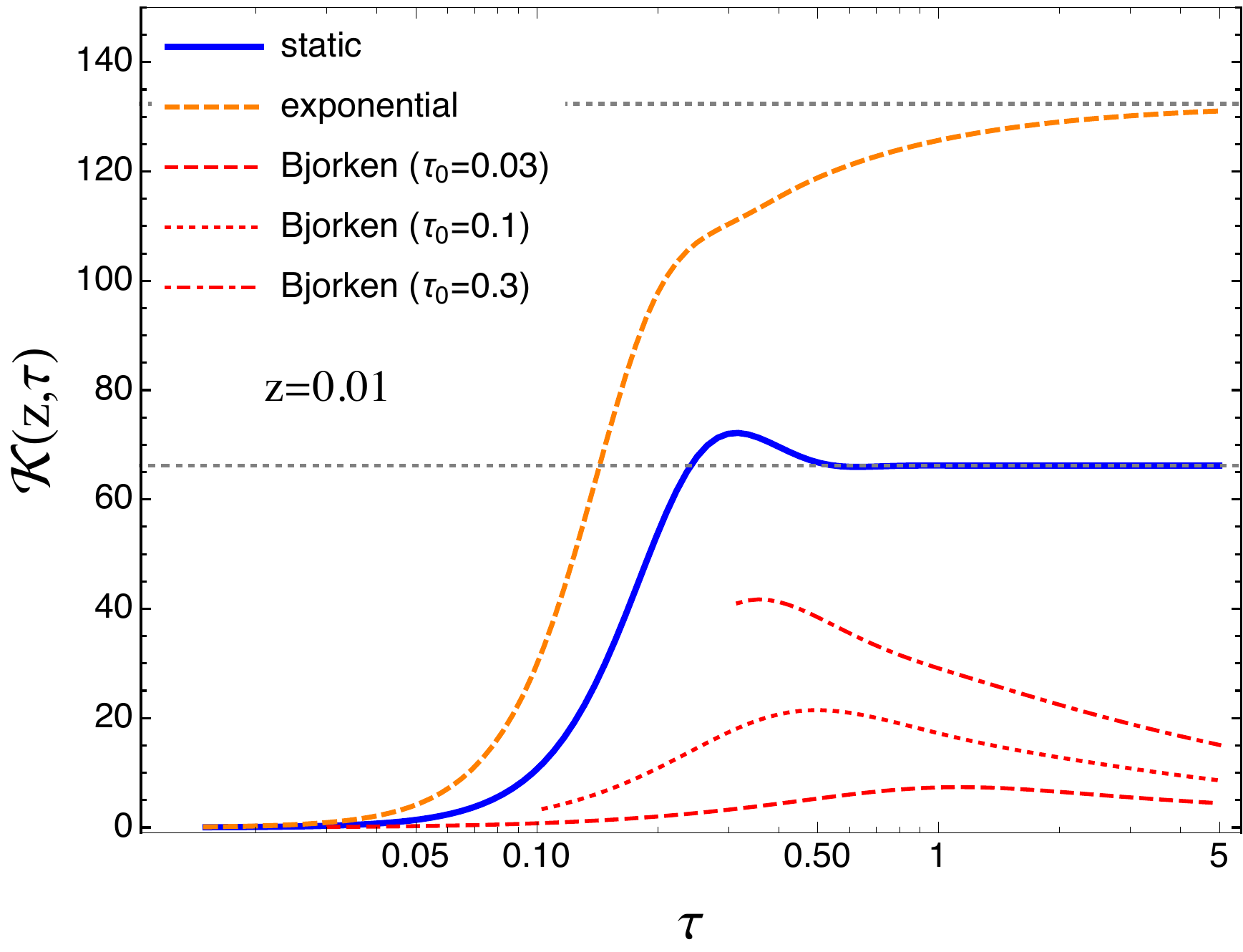}\;\;
\includegraphics[width=0.48\textwidth]{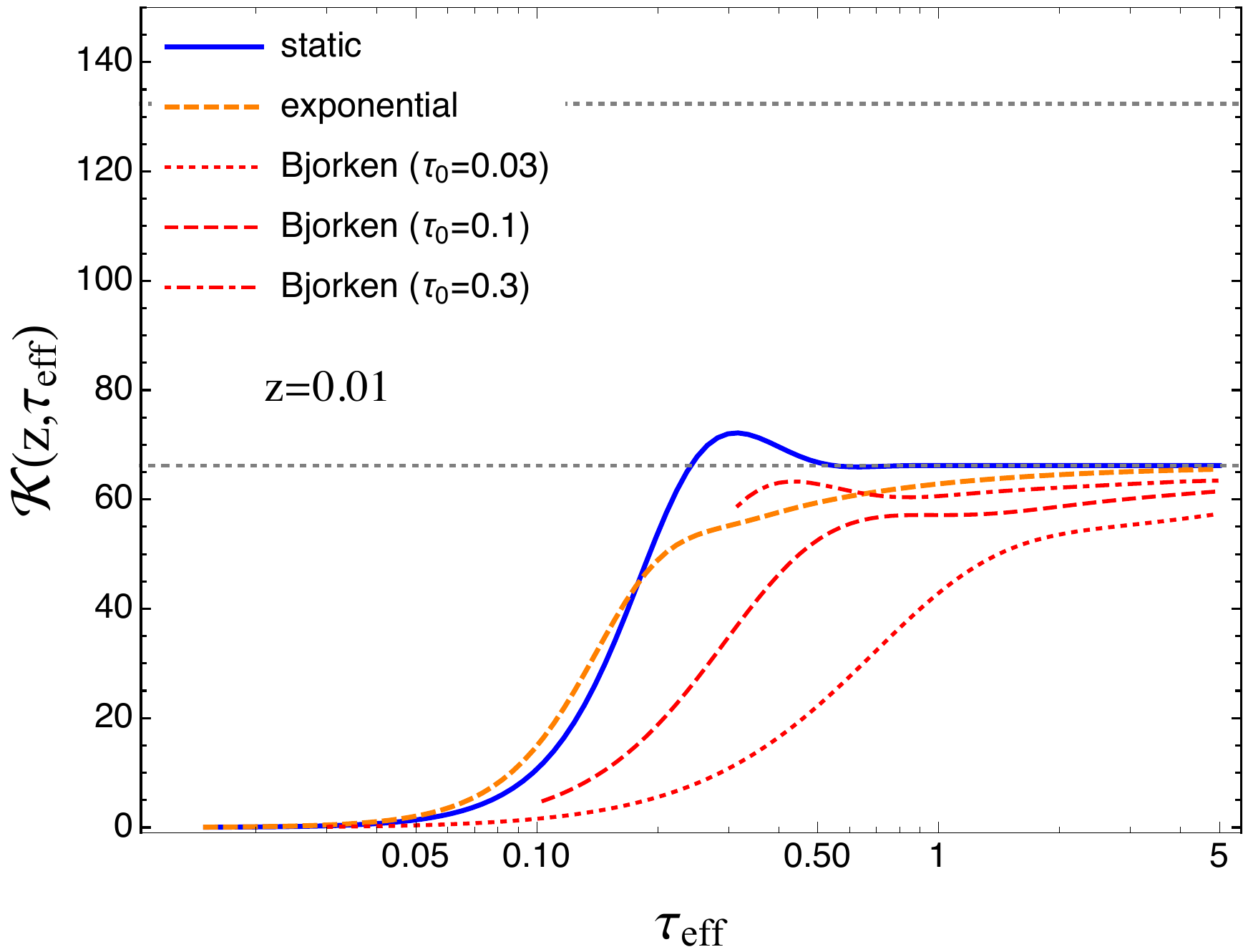}
\caption{Gluon emission rate for $z=0.01$ as a function of the evolution time $\tau$ (left) and $\tau_\text{eff}$ (right) for different medium expansion profiles. We plot the Bjorken case for three values of $\tau_0$ (the rate is zero for $\tau < \tau_0$). The plotting options are the same as in \autoref{fig:qhat}. The dotted lines are the asymptotic limits for the static, soft approximation, resulting from \autoref{eq:spectrum-static-soft}, and additionally scaled by a factor 2 to reproduce the long-time limit of \autoref{eq:rate-expo}.}
\label{fig:rate-tau}
\end{figure}
Turning next to the rate of medium-induced gluons, in the left panel of \autoref{fig:rate-tau}, we compare the resulting rates $\mathcal{K}(z,\tau)$ (${\cal K}(z,\tau,\tau_0)$ for the Bjorken model) at fixed $z=0.01$
plotted as a function of $\tau$ (left panel) and $\tau_\eff$ (right panel). The splitting rate for the static medium in the soft limit is constant, see (grey) dotted curves. For the exponential and full static cases, the splitting rate starts to grow from zero at $\tau = 0$, then it saturates at $\tau \sim \sqrt{z}$.  For the exponentially decaying medium, the rate saturates at slightly larger times compared to the static case, which is a consequence of limiting behavior of the ratio of Bessel's function in \autoref{eq:rate-expo}. In contrast, the rate for Bjorken expanding medium converges to zero at large times. While the ratio of Bessel's function in \autoref{eq:rate:bjorken} tends to one for $\tau \sim z /\tau_0$, the presence of the factor $\sqrt{\tau_0/(\tau_0+\tau)}$ leads to the dumping of the splitting rate for $\tau >\tau_0$.
%
The values of $\mathcal{K}(z,\tau)$ at large $\tau$ are larger for the exponential case than for the static case due to a longer effective extent of the medium, see definition \autoref{eq:definition-exp}. 

As discussed above, this mismatch can in principle, at least for soft sector, be corrected by the proper redefinition of the evolution time.
Summarizing the results in the previous section, the effective evolution time reads
\beq
\tau_\eff = \begin{cases}  \tau & \text{static medium} \\ 2 \tau  & \text{exponential medium} \\ 2 \sqrt{\tau_0 \tau} & \text{Bjorken medium} \end{cases} \,,
\eeq
where, we repeat, $\tau= \sqrt{\hat q/p} \, t$ in terms of the ``real'' in-medium distance $t$.
We plot the rate corresponding to the effective evolution time ${\cal K}(z,\tau_\eff)$ in the right panel of \autoref{fig:rate-tau}. While the resulting, effective rate for the exponential case scales closely with the static case, the Bjorken case depends strongly on the chosen value of the initial time $\tau_0$. Indeed, the scaling properties we derived only hold for sufficiently small momentum fractions $z$, also see below.

\begin{figure}
\centering
\includegraphics[width=0.48\textwidth]{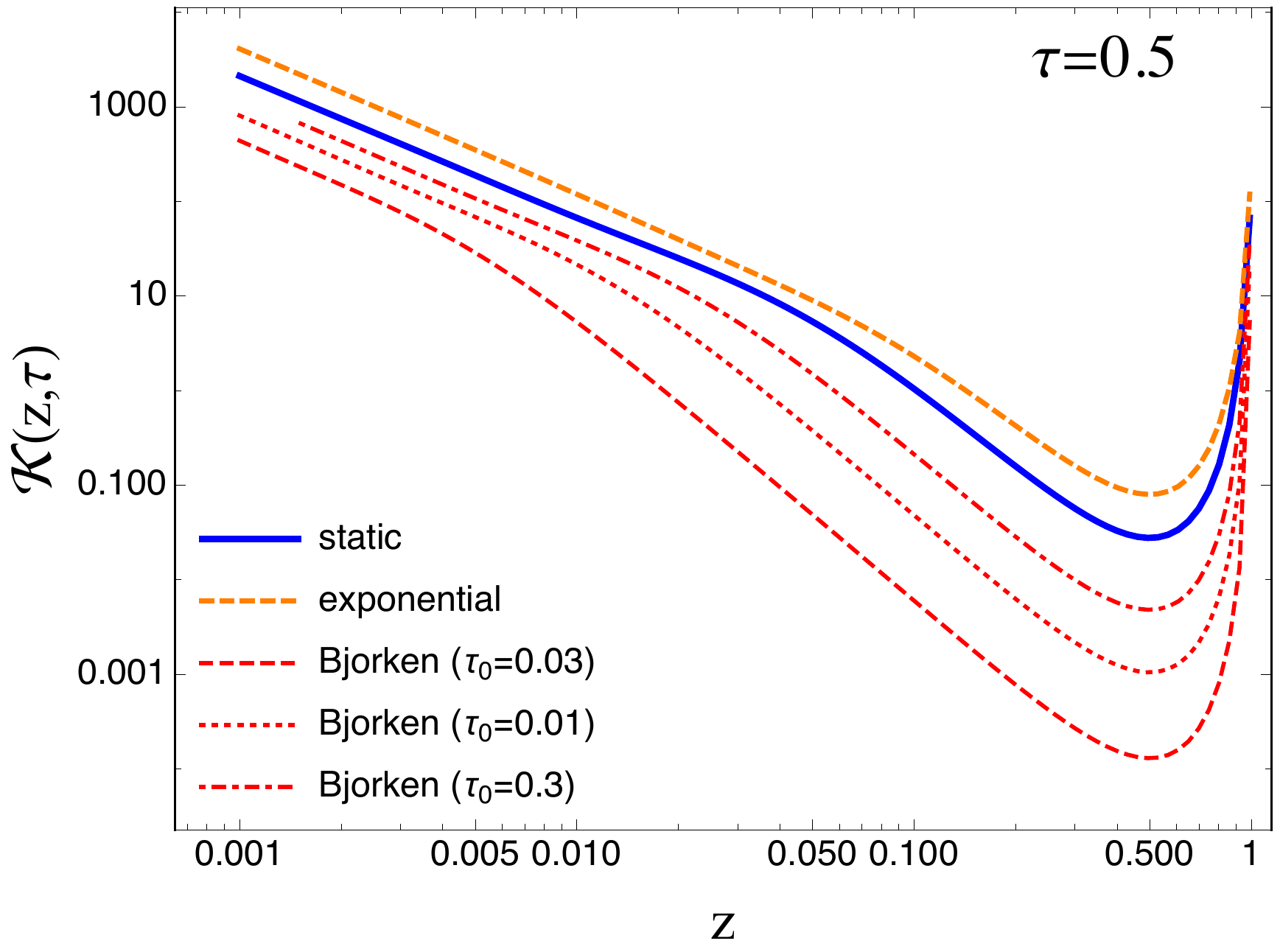}\;\;
\includegraphics[width=0.48\textwidth]{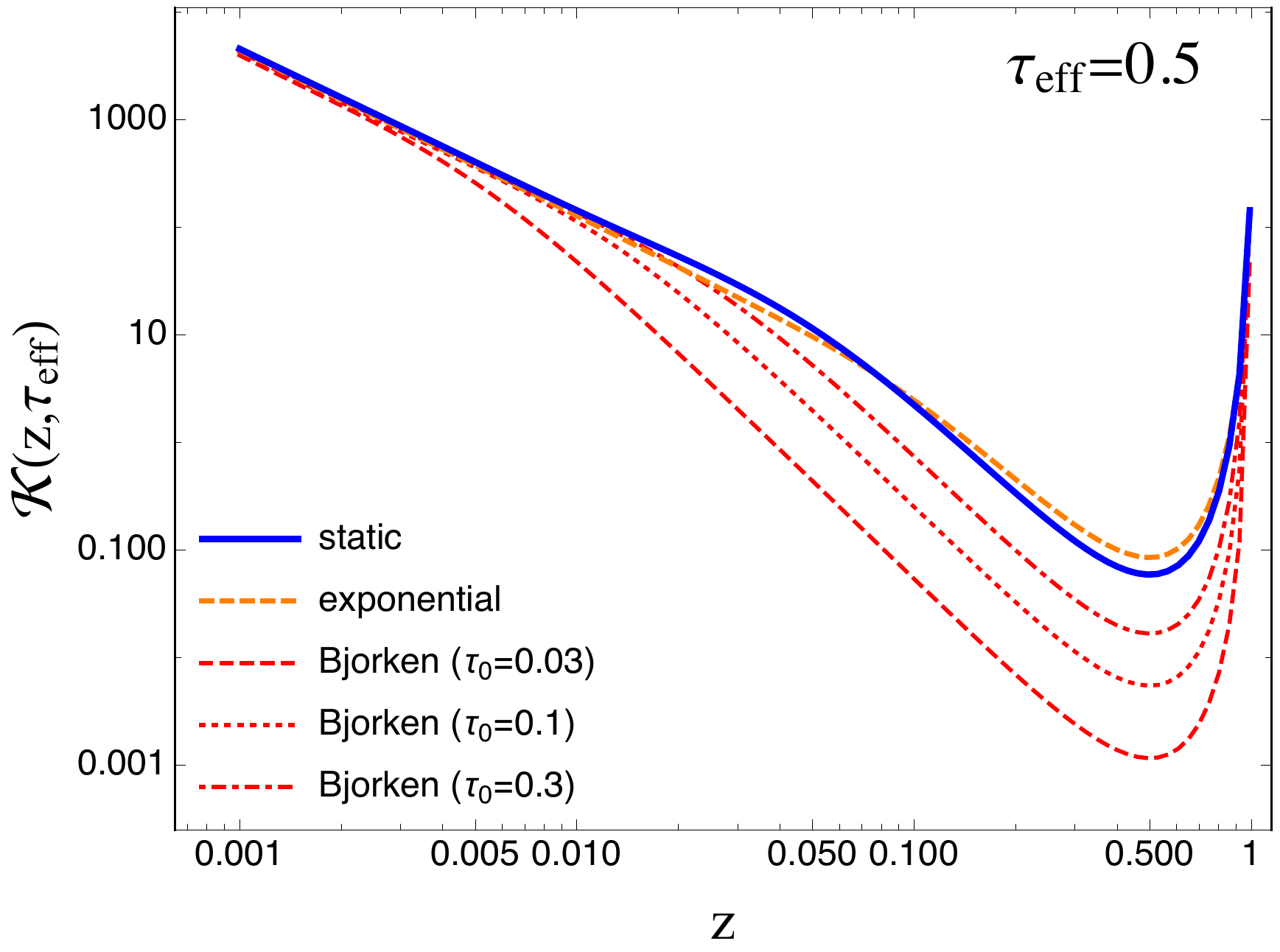}
\caption{Gluon emission rate for fixed  $\tau = 0.5$ (left) and $\tau_\eff=0.5$ (right) as a function of the momentum fraction $z$. The plotting options are the same as in \autoref{fig:qhat}.}
\label{fig:rate-z}
\end{figure}
This point becomes even more clear in \autoref{fig:rate-z} where we show a comparison of $\mathcal{K}(z,\tau)$ (left) and the corresponding ${\cal K}(z,\tau_\eff)$ (right) for fixed $\tau =0.5$ and $\tau_\eff = 0.5$, respectively, plotted as a function of the momentum fraction $z$. We observe in the left panel that, while for high values of $z$ the rates for different profiles differ significantly, the low-$z$ values they all have the same, universal slope which is a consequence of the $P(z) \kappa(z)$ factor present in splitting rates of all the profiles which diverges for $z \rightarrow 0$ as $z^{-3/2}$. We therefore expect to recover a universal behavior of the resulting parton branching evolution for expanding media in the soft gluon regime. The rate for fixed effective evolution time in the right panel confirms to a large extent this expectation for the exponential case. But for the Bjorken case, the scaling only holds for the soft sector, as expected from the scaling properties derived for the regime $z \ll \tau_0 \tau$ in \autoref{subsec:bjorken}.

\section{Rate equation for expanding medium}
\label{sec:rate-eq}

Equipped with the rate of emissions, we can now turn to the task of resumming multiple gluon emissions in the medium. In a large medium, possible interference terms are suppressed \cite{Blaizot:2012fh,Apolinario:2014csa}, and the resummation is performed via a kinetic rate equation. The evolution equation for the energy distribution of medium-induced gluons, $D(x,\tau) = x\, \dd N/\dd x$, is given by \cite{Blaizot:2013hx,Blaizot:2013vha}
\begin{align}
\label{eq:RateEquation-generic}
\frac{\partial D(x, \tau)}{\partial \tau} &= \int_0^1 \dd z \,\mathcal{ K}(z,\tau) \left[\sqrt{\frac{z}{x}} D\left(\frac{x}{z},\tau \right) \Theta(z-x) - \frac{z}{\sqrt{x}} D(x,\tau) \right] \,.
\end{align}
The initial value of the $D(x,\tau)$ is a $\delta$-function at $x=1$ which characterizes the initial single color charge entering the evolving medium. Furthermore, conservation of energy implies
\beq
\int_0^1 \dd x\, D(x,\tau) = 1 \,. 
\eeq
While this is formally violated due to the soft singularity at $x=0$, this can be reinstated by assuming the accumulation of energy at the thermal scale $x \sim T/p$ where elastic re-scattering leads to thermalization \cite{Baier:2000sb}. For the Bjorken expansion scenario, the distribution also depends on the initial time $\tau_0$, i.e. $D(x,\tau,\tau_0)$. In this Section, we have fixed $\tau_0 = 0.03$.

\begin{figure}
\centering
\includegraphics[width=0.5\textwidth]{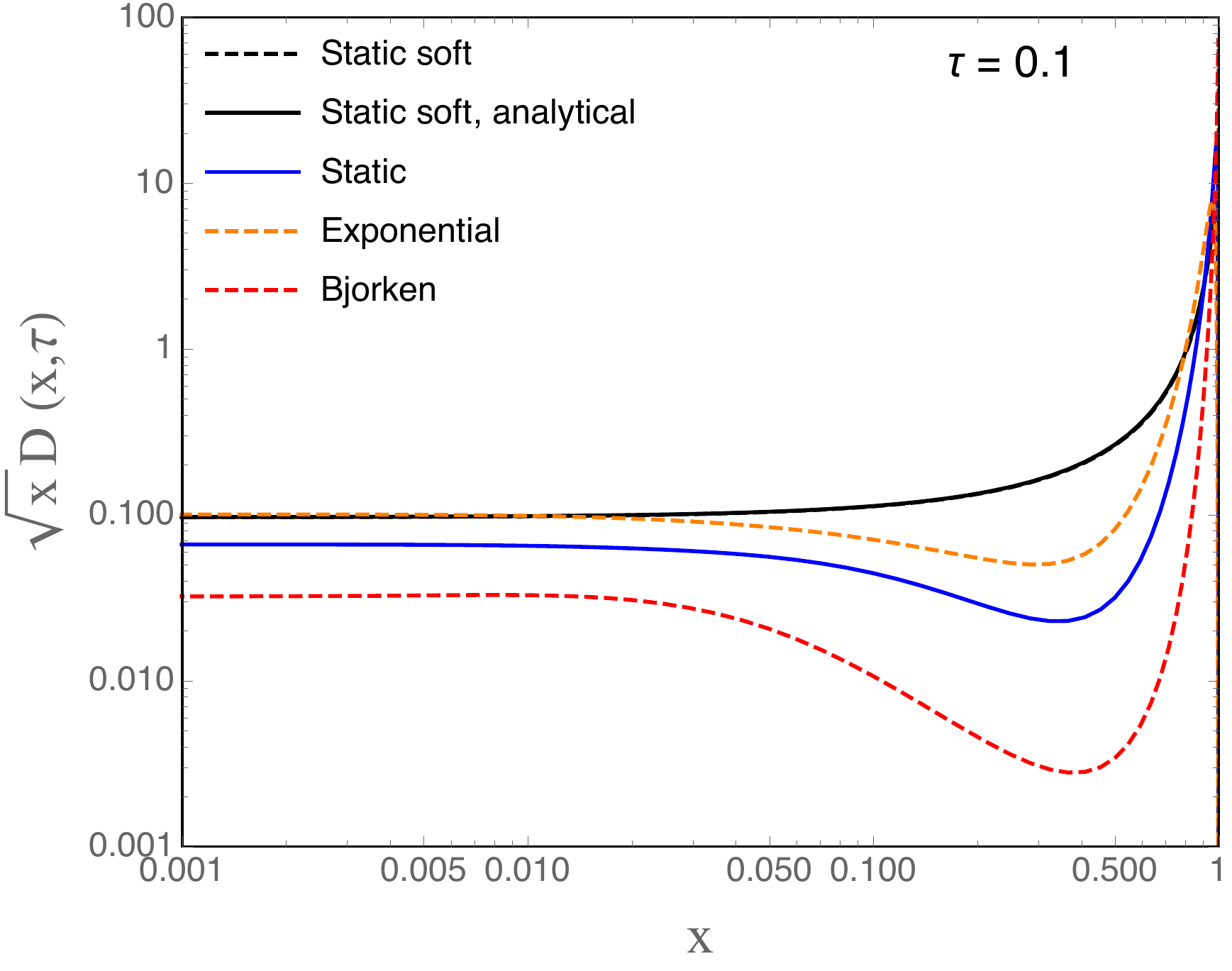}%
\includegraphics[width=0.5\textwidth]{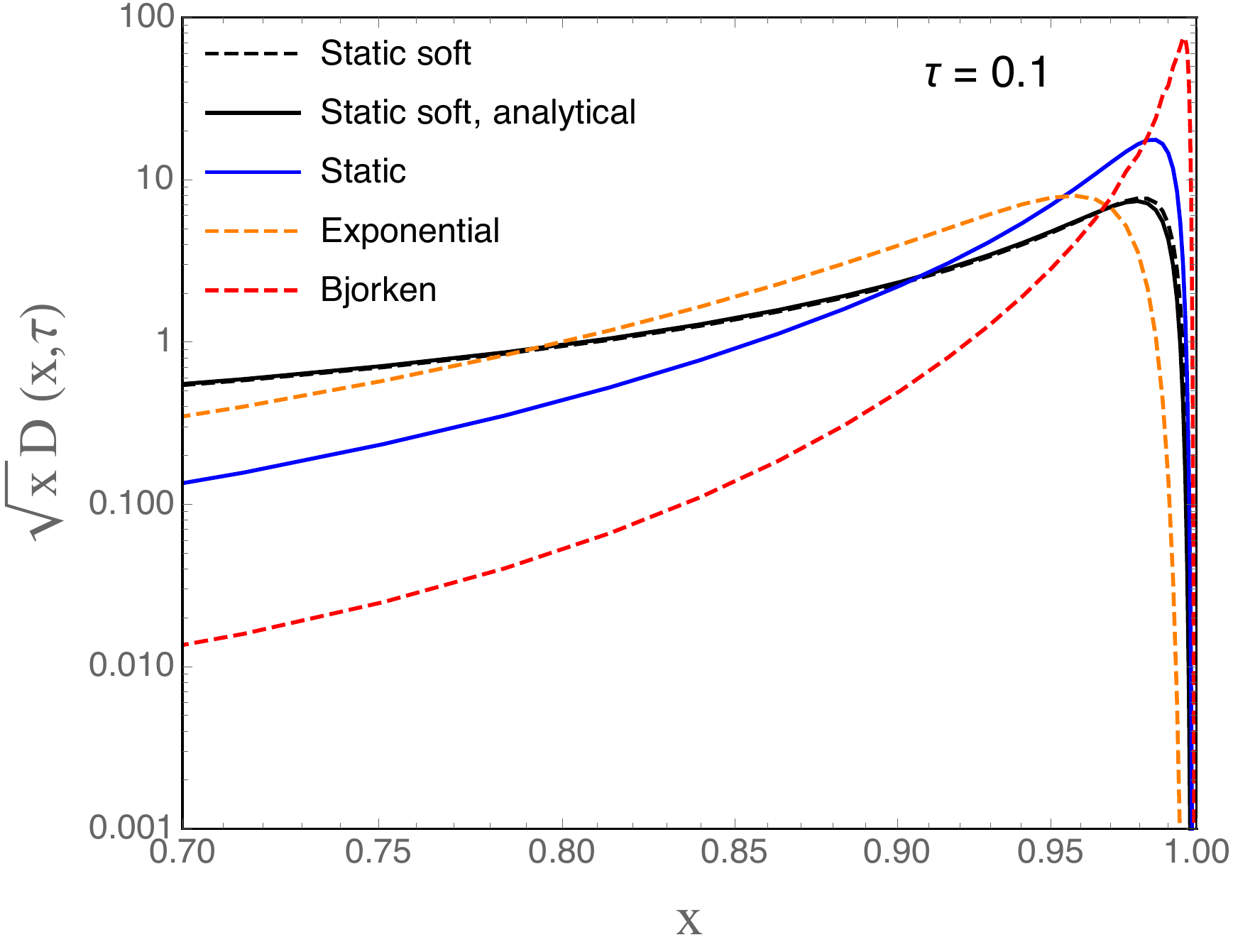}
\centering
\includegraphics[width=0.5\textwidth]{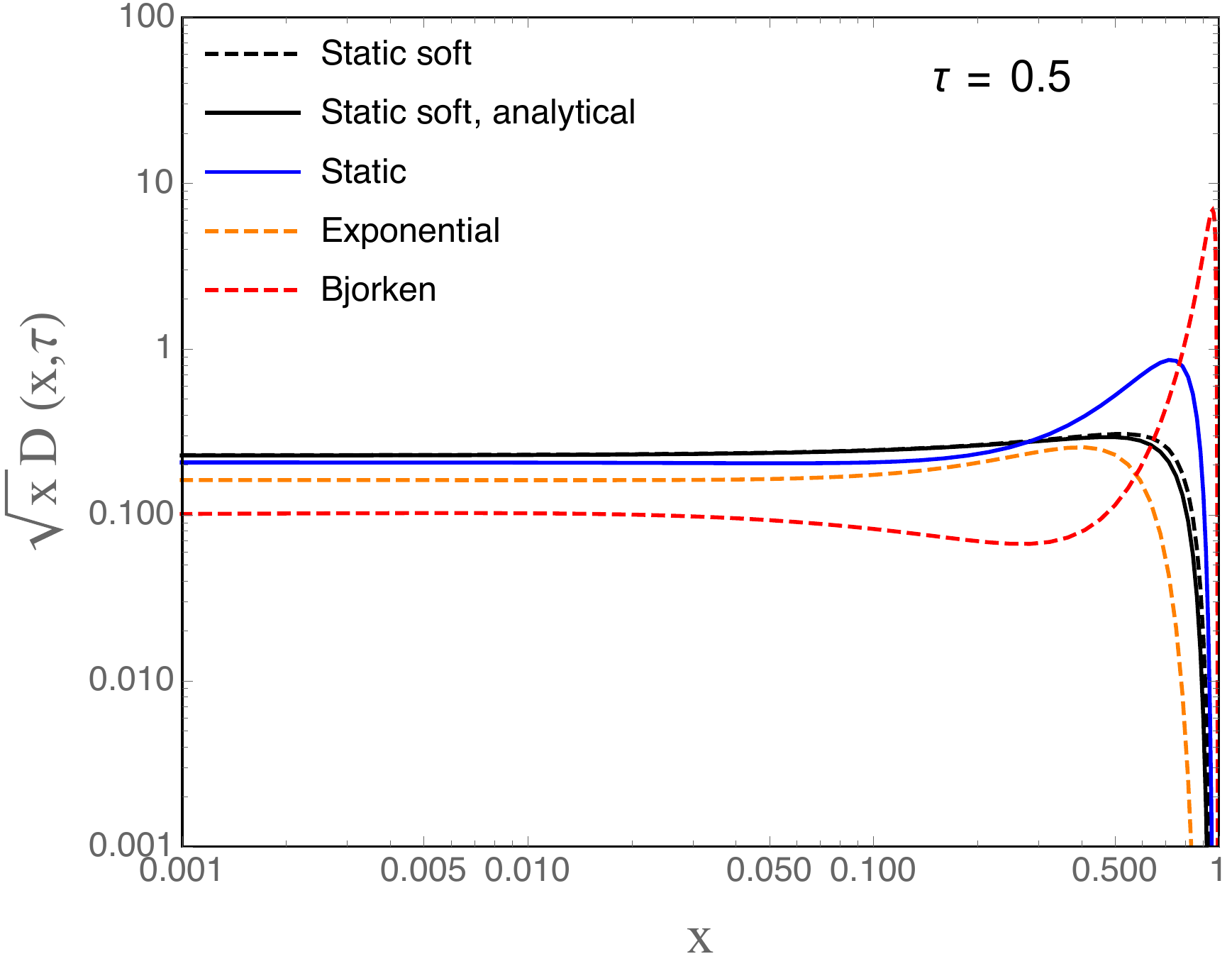}%
\includegraphics[width=0.5\textwidth]{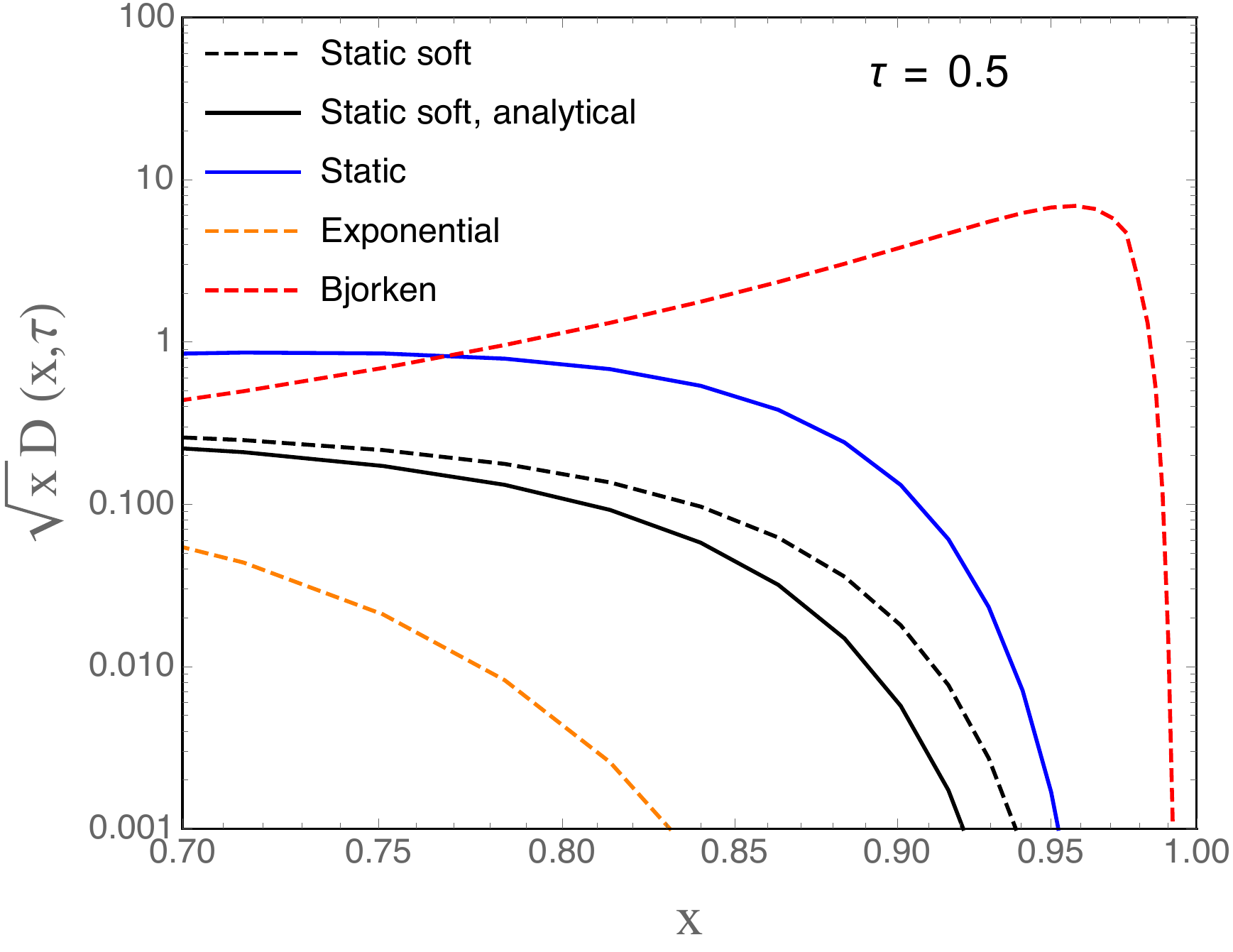}
\centering
\includegraphics[width=0.5\textwidth]{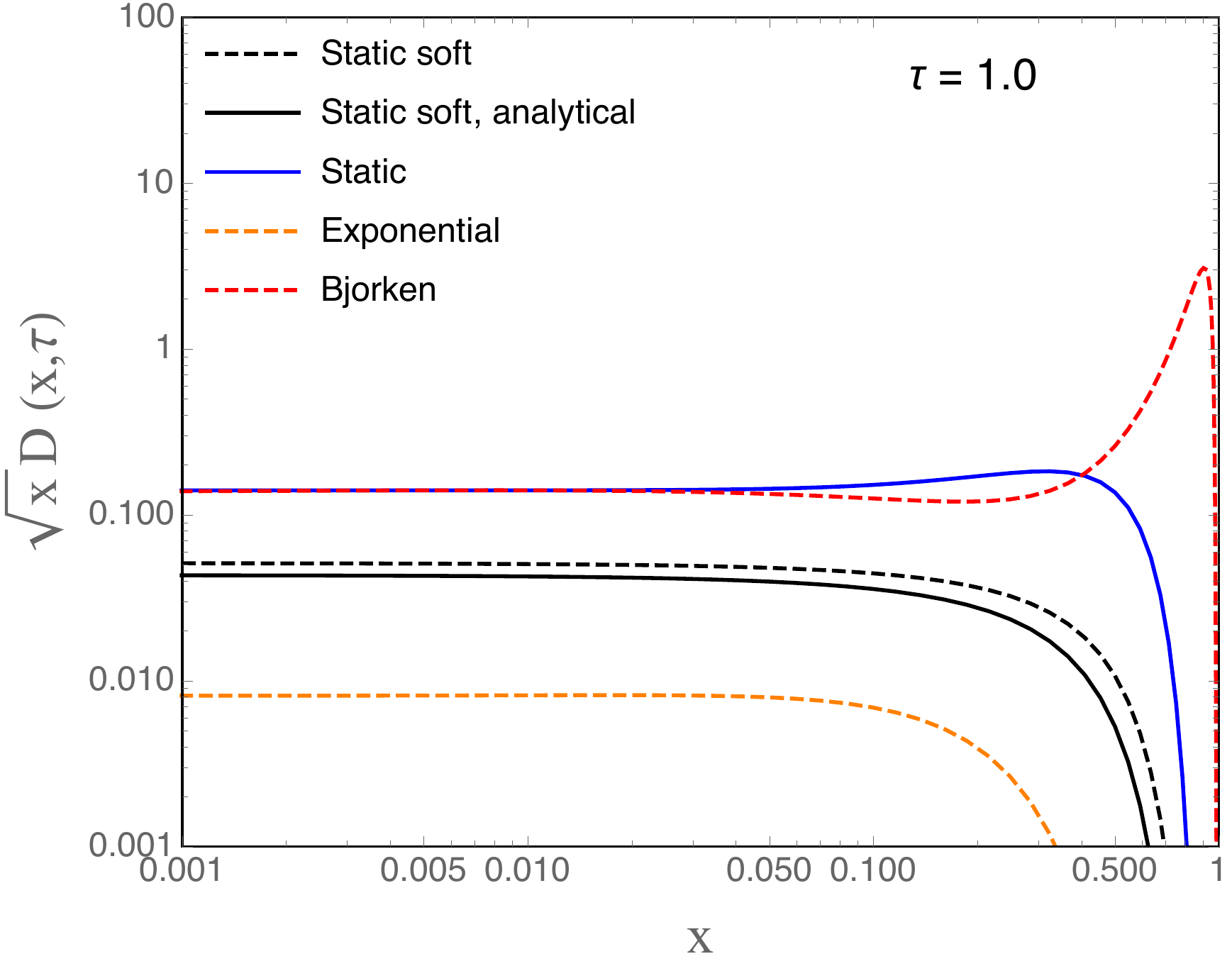}%
\includegraphics[width=0.5\textwidth]{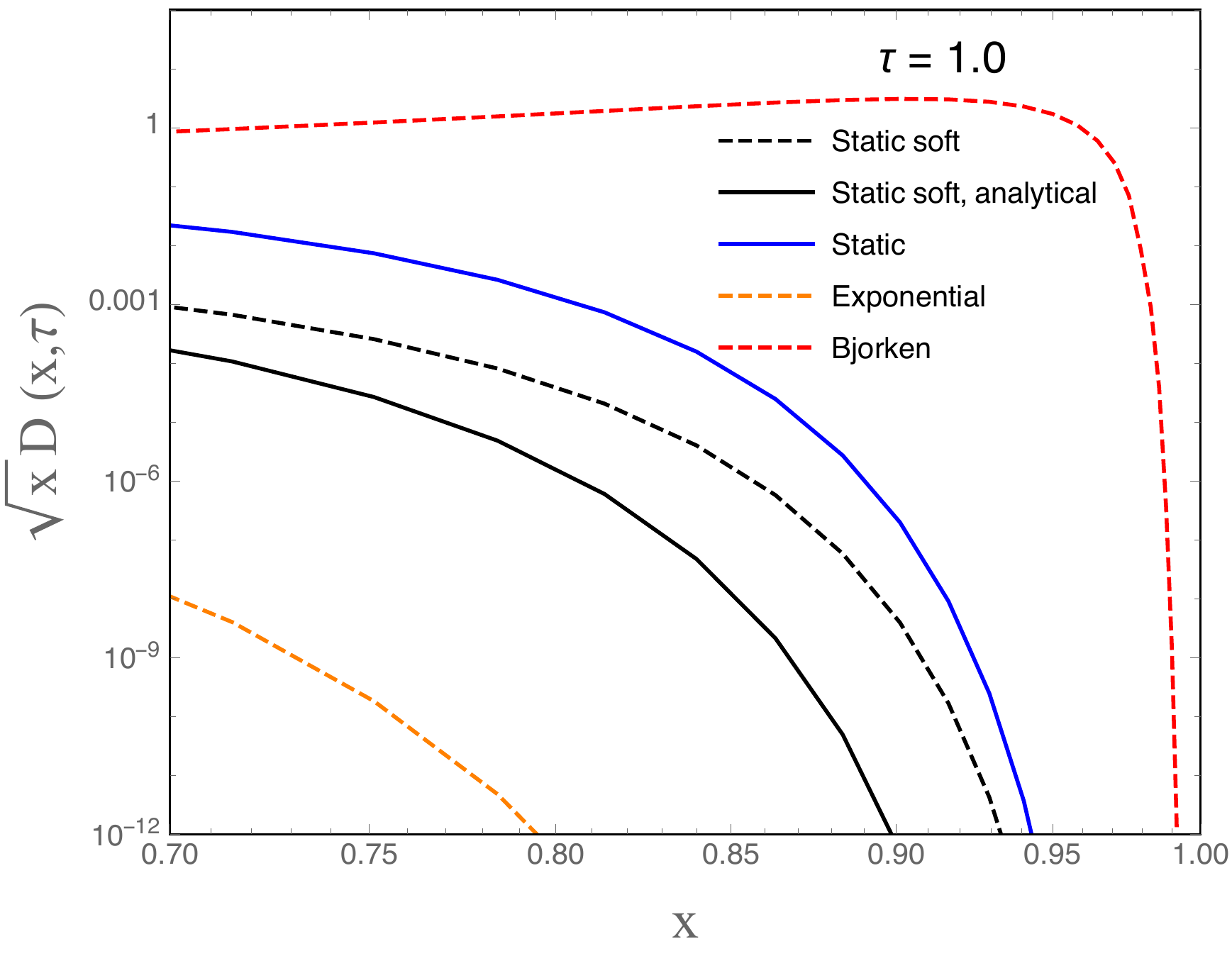}
\caption{
Medium induced gluon distribution $\sqrt{x}D(x,\tau)$ for three different values of $\tau$ and three types of medium expansion calculated numerically: static soft (dashed black), static (solid blue), exponential (dashed orange), and Bjorken (dashed-dotted red). We also plot the soft limit of the static medium calculated analytically (solid black) for reference.
Left panels show the full distribution, right panels zoom in the high-$x$ region.
 }
\label{fig:distributions-allx}
\end{figure}
\begin{figure}
\centering
\includegraphics[width=0.5\textwidth]{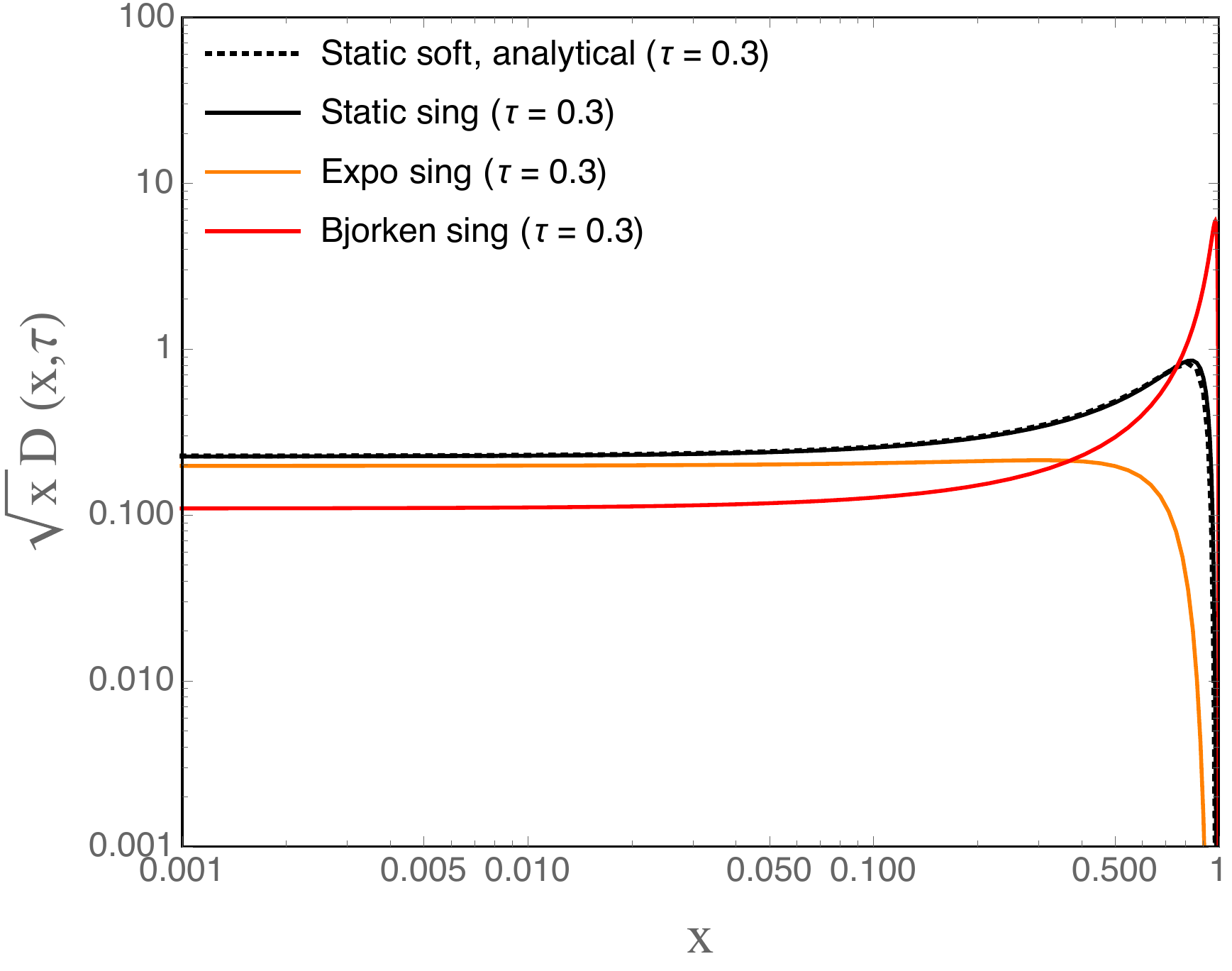}%
\includegraphics[width=0.5\textwidth]{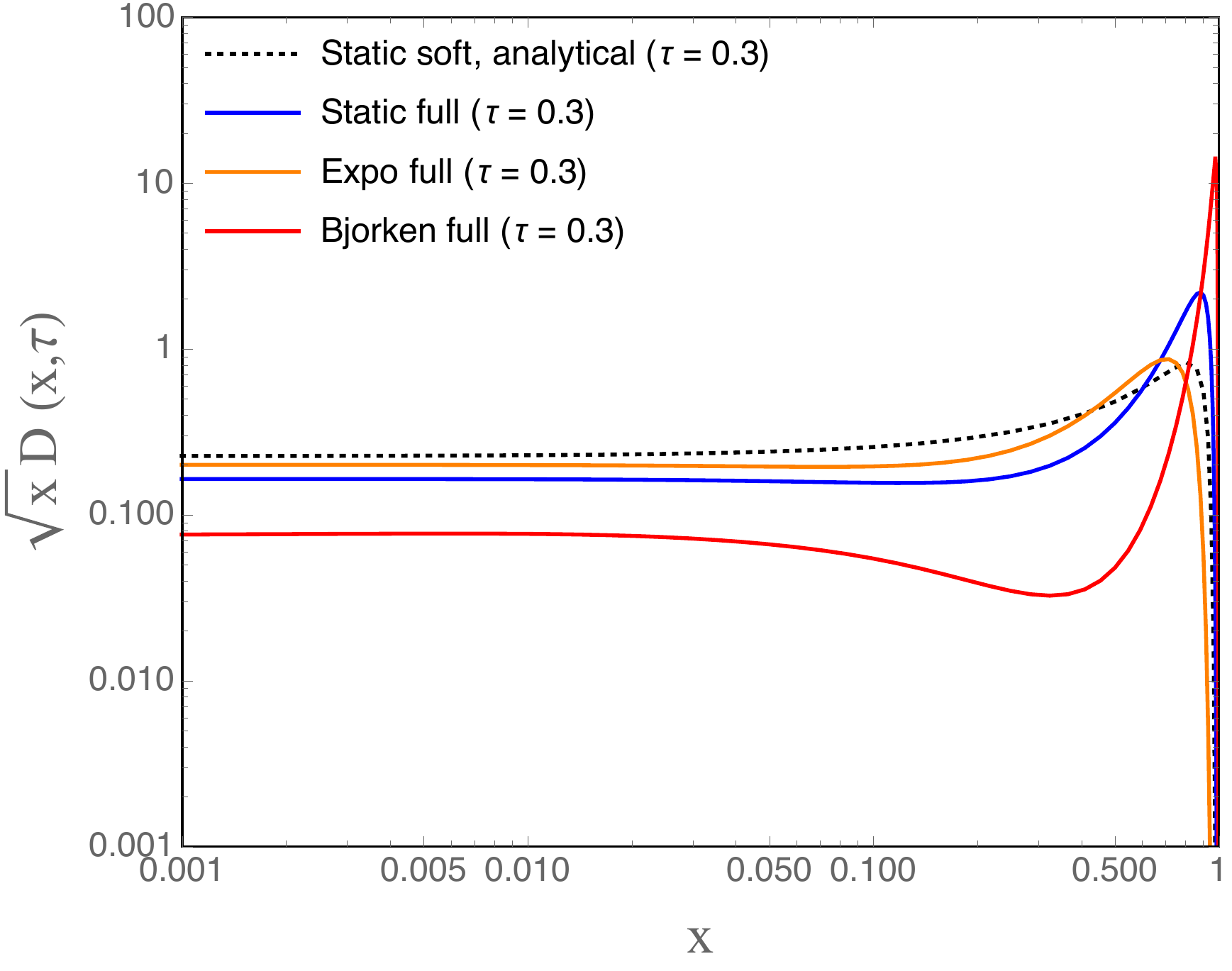}
\centering
\includegraphics[width=0.5\textwidth]{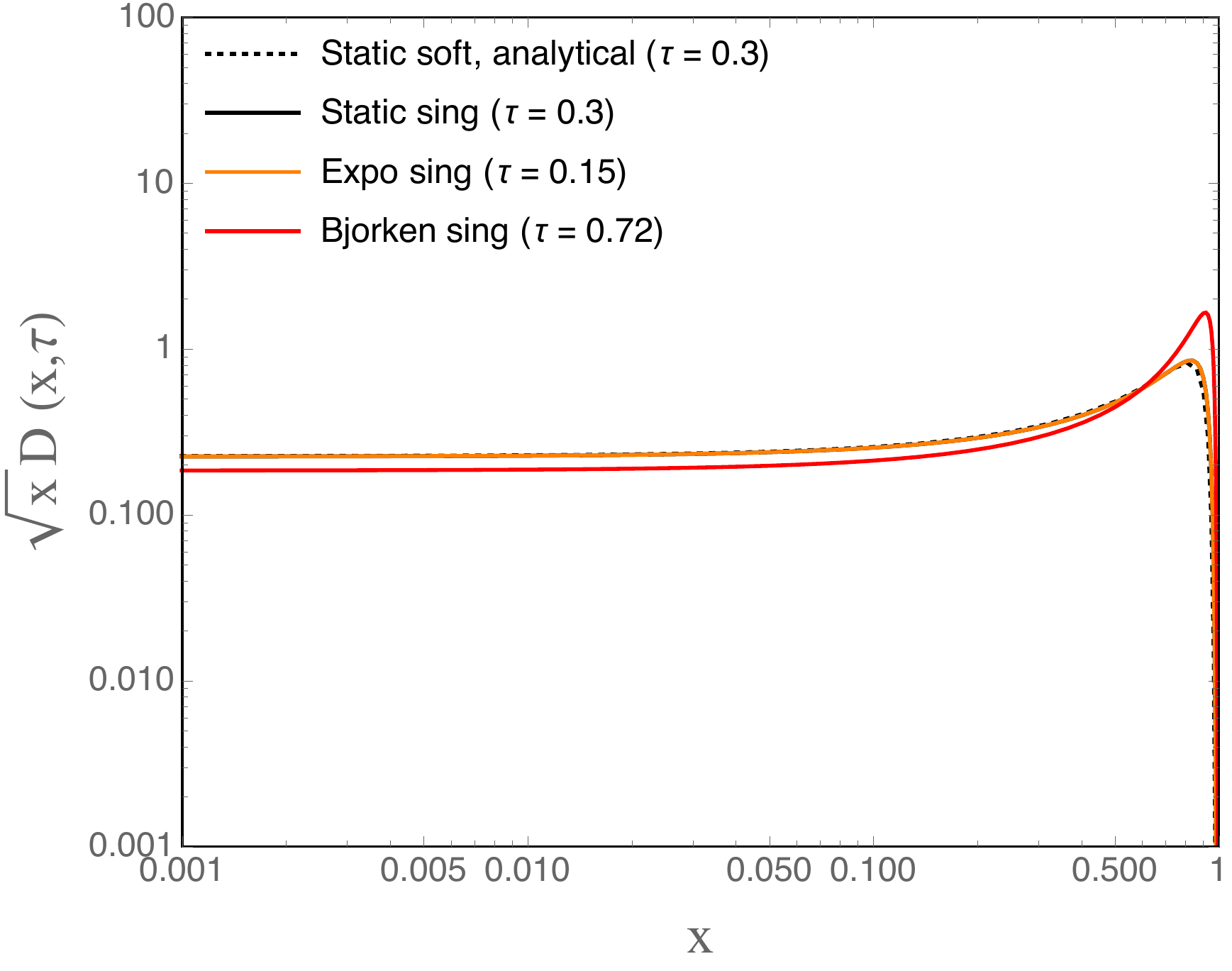}%
\includegraphics[width=0.5\textwidth]{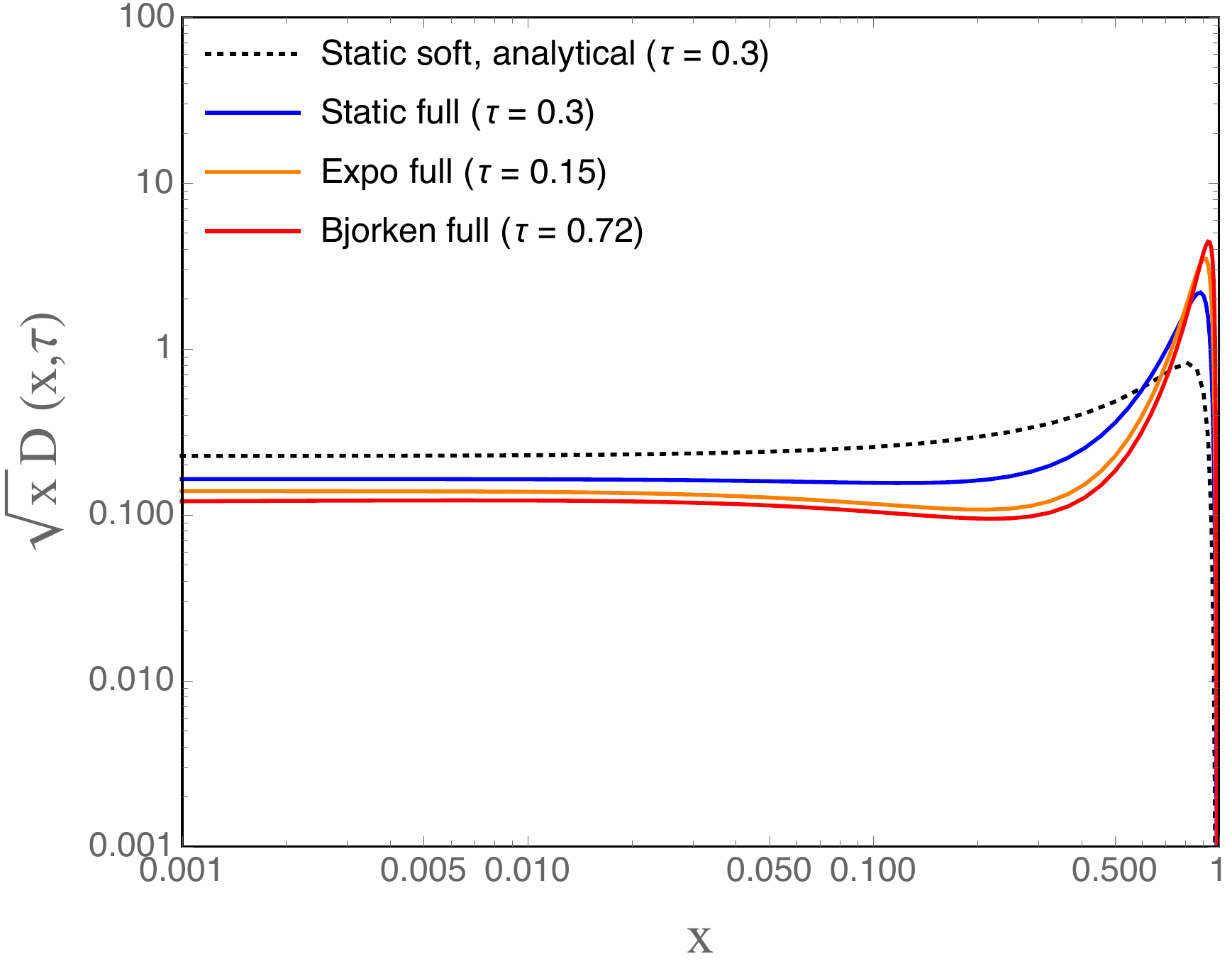}
\caption{
Medium induced gluon distribution $\sqrt{x}D(x,\tau)$ for singular rates (left panels) and full rates (right panels). The upper row corresponds to the distributions for different medium expansions evaluated at equal evolution time $\tau$. In the lower row, we have plotted the distributions at a fixed \emph{effective} evolution time $\tau_\eff$.}
\label{fig:distributions-scaling}
\end{figure}

The rate equation \autoref{eq:RateEquation-generic} was solved numerically for the static medium, exponentially decaying medium, and the Bjorken case introduced in 
\autoref{sec:spectrum}. 
The resulting distributions of $D(x,\tau)$ are shown for three representative values of $\tau$ in \autoref{fig:distributions-allx}, see figure caption for further details. Despite the differences observed in the rates, at low $x$, all the $D(x)$ distributions converge to a universal scaling with $1/\sqrt{x}$ which is a consequence of the low-$x$ behavior discussed in \autoref{sec:rate_properties} driven by a presence of factors $P(z) \kappa(z)$ in all the splitting rates. The magnitude of the effects is different and is expected to scale with the average parameter $\langle \hat q \rangle$ which is hierarchically $\langle \hat q \rangle_{\rm exp} > \langle \hat q \rangle_{\rm static} > \langle \hat q \rangle_{\rm Bjork}$ (for the choice of parameters used here).
At high-$x$, which predominantly drives the jet suppression factor, see \autoref{sec:moments}, the reduction of $D(x,\tau)$ is the strongest for the exponentially decaying medium and the weakest for the Bjorken case, according to the established hierarchy of $\langle \hat q \rangle$ for the parameters used here. 

As a check of our numerical routine, we have also evaluated the distribution for the static soft limit, i.e. where we use \autoref{eq:rate-static-soft} as the splitting rate. These results can be compared with the known analytical solution \cite{Blaizot:2013hx},
\beq
\label{eq:rate-equation-solution}
D_{\rm sing}(x,\tau) = \frac{\bar \alpha \tau}{\sqrt{x}(1-x)^{3/2}} \text{e}^{-\pi \frac{\bar \alpha^2 \tau^2}{1-x}} \,,
\eeq
where the sub-script refers to the ``singular'' rate in \autoref{eq:rate-static-soft}, see also \cite{Blaizot:2015jea}.
The numerical and analytical results are plotted in \autoref{fig:distributions-allx} as the solid (black) and dashed (black) curves and we see a good agreement over a wide range in $x$ and $\tau$. In this situation, the energy stored in the spectrum decreases exponentially with $\tau$, ${\cal E}(\tau) = \int_0^1 \dd x\, D(x,\tau) = \rme^{-\pi \bar \alpha^2\tau^2}$ \cite{Blaizot:2013hx}.\footnote{Note that the authors of Ref. \cite{Blaizot:2013hx} use directly $\tilde\tau=\bar\alpha \tau$ as the evolution time.} 

Finally, in order to confirm the scaling properties uncovered in the last Section, in \autoref{fig:distributions-scaling} we compare the resulting gluon distribution $D(x,\tau)$ using only the singular rates, i.e. \autoref{eq:rate-static-soft} for the static case, \autoref{eq:rate-expo-approx} for the exponential case and \autoref{eq:rate-Bjork-approx} for the Bjorken case, to the distributions obtained with the full rates. In the upper panels, the distributions are plotted for a fixed, common evolution time $\tau$ while below the distributions are evaluated at fixed $\tau_\eff$. The distributions resulting from the singular rates in the left panels clearly show nice scaling properties. The full distributions, plotted in the right panels of \autoref{fig:distributions-scaling}, do not respect a scaling with $\tau_\eff$.

As we will illustrate next on the level of the jet suppression factor, a better scaling is actually achieved by using the average jet quenching parameter $\langle \hat q \rangle$, as defined in Eqs.~\ref{eq:average-qhat} and \ref{eq:average-qhat-2}. However, it is clear from \autoref{fig:distributions-allx} and \autoref{fig:distributions-scaling} that the distributions of medium-induced gluons produced in differently expanding media have different features at both low and high values of $x$. 

\section{Moments of $D(x,\tau)$ and the jet suppression factor}
\label{sec:moments}
One of the key observables quantifying inclusive jet suppression is the jet nuclear modification factor, measured, for instance, by the LHC experiments \cite{Aad:2014bxa,Aaboud:2018twu,Khachatryan:2016jfl,Acharya:2019jyg}. The yield for the inclusive jet suppression can be obtained as a convolution of the $D(x,\tau)$ distribution with the initial parton spectra,
\beq
\frac{\dd \sigma_{\rm AA}}{\dd \pT} = \int_0^\infty \dd \pT' \int_0^1 \dd x \,\delta(\pT- x \pT') D\left(x, \tau \equiv \sqrt{\hat q_0 / \pT'}L \right) 
\frac{\dd \sigma_0}{\dd \pT'} \,,
\eeq
see e.g. \cite{Baier:2001yt,Salgado:2003gb,Mehtar-Tani:2014yea}.
Note that the evolution time $\tau$ now depends on the unknown initial energy of the parton.
The initial parton spectra can be approximated by a power law, $\dd \sigma_0/\dd \pT \propto \pT^{-n}$. In this case, the jet suppression factor $\Raa(\pT) = \big(\dd \sigma_{\rm AA}/\dd \pT\big) \big/ \big(\dd \sigma_0/\dd \pT\big)$, is 
\beq
\label{eq:suppression-factor-1}
\Raa(\pT) = \int_0^1 \dd x \, x^{n-1} D(x, \sqrt{x} \tau) \,,
\eeq
where now $\tau = \sqrt{\hat q_0/\pT}L$, as before. For the Bjorken model, the distribution has additionally a dependence on the initial time $\tau_0$, and the distribution in the integrand in \autoref{eq:suppression-factor-1} becomes $D(x,\sqrt{x}\tau,\sqrt{x}\tau_0)$.

Let us focus for a moment on the analytical solution of the rate equation given by \autoref{eq:rate-equation-solution}. At the present stage, it is illuminating to change variables to 
$\epsilon = \pT(1-x) $, 
where $\epsilon$ has the meaning of the energy lost by the particle due to medium-induced emissions. In this case,
\begin{align}
\label{eq:raa-analytic-1}
\Raa(\pT) &= \int_0^{\pT} \dd \epsilon \, \left(1- \frac{\epsilon}{\pT} \right)^{n-1} \sqrt{\frac{\omega_s}{\epsilon^3}} \rme^{- \frac{\pi\omega}{\epsilon} \left(1-\frac{\epsilon}{\pT} \right)} \,,
\end{align}
where $\omega_s = \bar\alpha^2 \hat q_0 L^2$ is the scale of soft, multiple gluon emissions.
In the limit of $\omega_c \ll \pT$, this expression can be approximated by
\beq
\label{eq:raa-analytic-2}
\Raa(\pT) \approx \int_0^{\infty} \dd \epsilon \, \rme^{-\nu \epsilon} \sqrt{\frac{\omega_s}{\epsilon^3}} \rme^{- \frac{\pi\omega}{\epsilon}} = \rme^{-2 \sqrt{\pi \omega_s \nu}}\,,
\eeq
where $\nu = (n-1)/\pT$. This is nothing else than the (inverse) Laplace transform of the energy loss distribution ${\cal P}(\epsilon) = \sqrt{\omega_s/\epsilon^3}\, \rme^{-\pi \omega_s/\epsilon}$. This quantity is normalized $\int_0^\infty \dd \epsilon\, {\cal P}(\epsilon) = 1$.
We have checked numerically that the approximation in going from \autoref{eq:raa-analytic-1} to \ref{eq:raa-analytic-2} is valid to a few percent over a large range of $\pT$.

We will now discuss the quenching factor for two analytically available limiting cases, namely i) the quenching weights that only account for primary medium-induced gluon emissions off the jet and ii) the expectations from the distribution $D_\text{sing}(x,\tau)$ of soft emissions \autoref{eq:rate-equation-solution}. Finally, we address the numerical results using the full rates.

\subsection{Jet suppression from primary emissions}
\label{sec:suppresion-qw}
Since the dominant contribution to the jet quenching factor comes from large $x$, or small $\epsilon/\pT$, one can consider a simpler scenario, where we neglect the further branching of the primary gluons that are emitted from the leading particle. Assuming independent emissions as before, this approximation scheme is typically referred to as the ``quenching weights'' \cite{Baier:2001yt,Salgado:2003gb}. It turns out that the Laplace transformed energy loss distribution in this case, simply reads \cite{Baier:2001yt}
\beq
\label{eq:quenching-weight-approx}
\tilde D(\nu) = \exp \left[ - \int_0^\infty \dd z\, \rme^{-z} N\left(\frac{z}{\nu} \right) \right] \,,
\eeq
where $N(\omega) = \int_\omega^\infty \dd \omega' \frac{\dd I}{\dd \omega'}$ is the multiplicity of gluons carrying energy $\omega$ and above. The resulting quenching factor is nothing else than $\Raa = \tilde D(n/\pT)$. The main contribution to quenching comes from the regime of multiple, soft gluon emissions with $\omega \sim \bar \alpha^2 \omega_c$.

For the static medium, Baier-Dokshitzer, Mueller and Schiff (BDMS) \cite{Baier:2001yt} found a compact  formula approximating the multiplicity,
\beq
\label{eq:static-multiplicity}
N_\text{stat}(u) = 2 \bar \alpha \left(\sqrt{\frac{2}{u}} - \log 2 \log \frac{1}{u} - 1.44135\ldots \right) \,,
\eeq
where $u \equiv \omega /\omega_c$. The two first terms of this expression are found from the soft gluon approximation, $\omega \dd I_\text{stat} /\dd \omega \approx 2 \bar \alpha \left( \sqrt{\omega_c/(2\omega)} - \log 2 \right)$, while the numerical factor is found by matching to the full spectrum. Note that a na\"ive application of the soft gluon limit would not contain the two last terms on the right hand side of \autoref{eq:static-multiplicity}, and would have resulted in an overestimation of the quenching. The approximation in \autoref{eq:static-multiplicity} works surprisingly well for realistic values of the medium parameters and results in a suppression factor that is almost exactly the one obtained by directly using the full BDMPS spectrum in \autoref{eq:quenching-weight-approx}. 

\begin{figure}[b]
\centering
\includegraphics[width=0.48\textwidth]{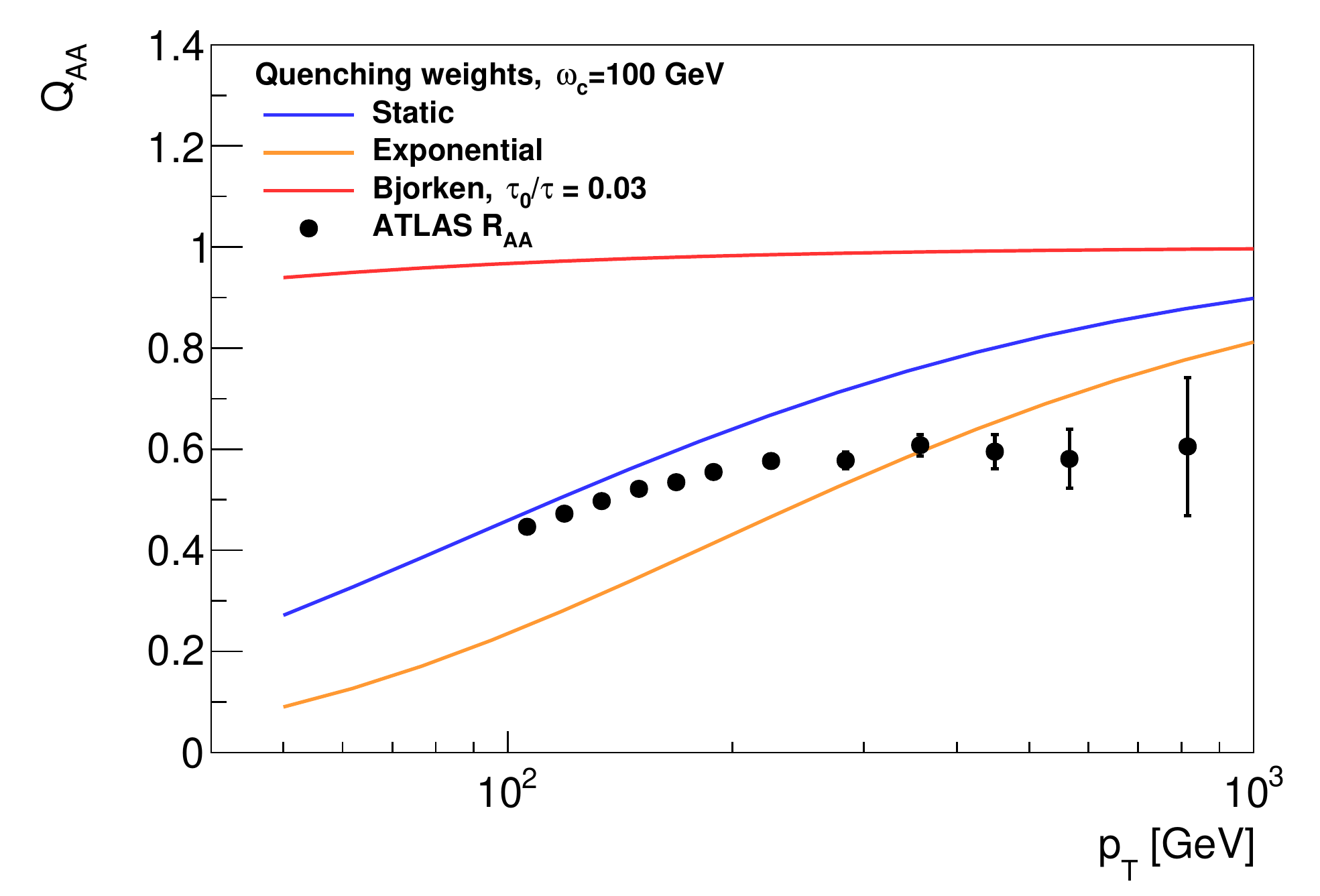}\;\;
\includegraphics[width=0.48\textwidth]{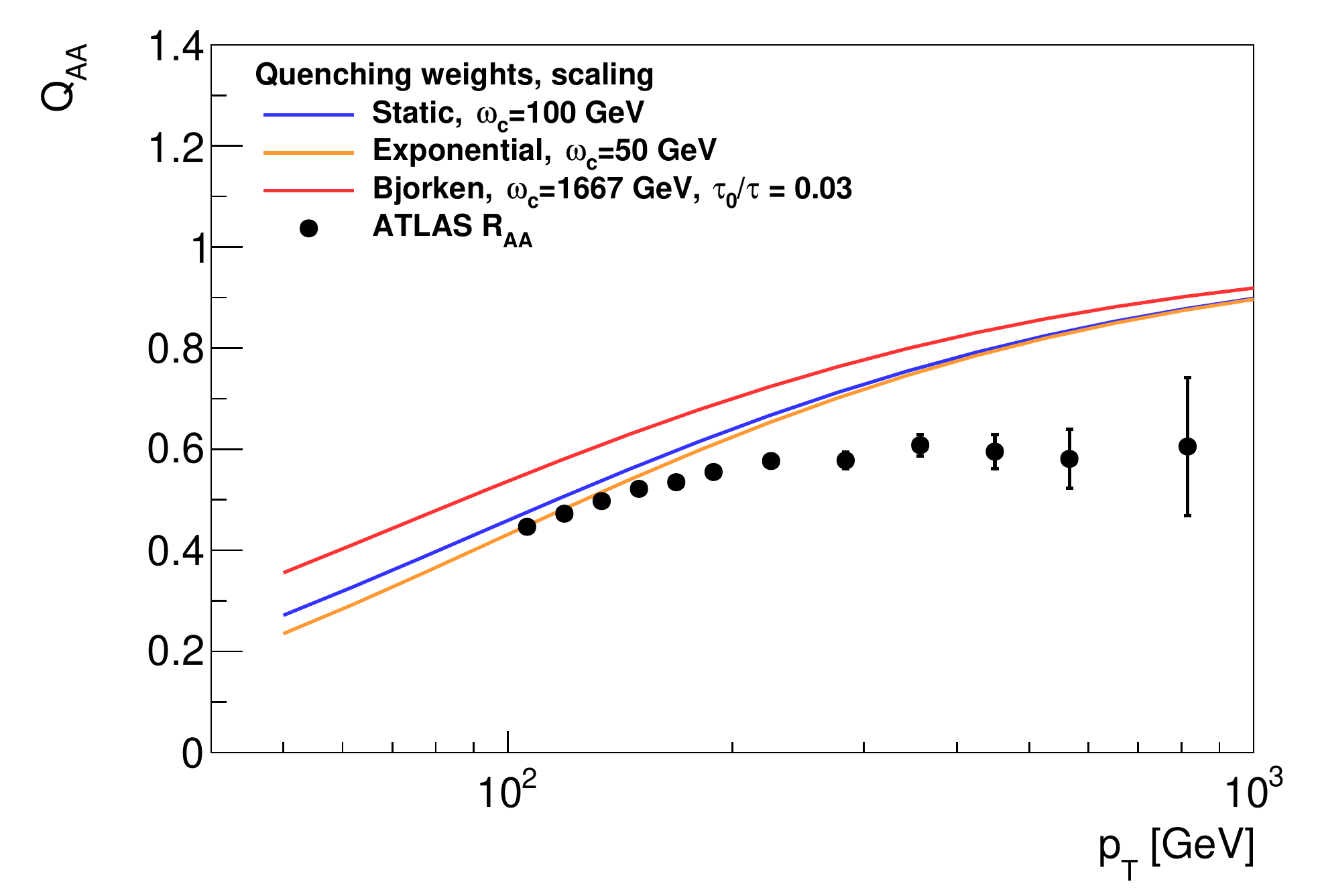}
\caption{Scaling properties of the jet suppression factor $\Raa$ for different medium expansions in the quenching weights approximation. In the left panel we plot the suppression given a fixed value of $\omega_c = \hat q_0L^2/2$; in the right panel we plot the suppression factors for a fixed $\langle \omega_c \rangle = \langle \hat q \rangle L^2/2$ for the different media.}
\label{fig:qw-scaling}
\end{figure}
Following a similar logic, we have not been able to find a similar compact approximation for the multiplicity in the exponentially and Bjorken expanding scenarios. However, one should expect a similar dependence on an effective ratio of the gluon energy $\omega$ and a maximal energy scale $\sim \omega_c$. Recalling the left panel of \autoref{fig:spectra-scaling}, it is therefore natural to expect that the optimal scaling parameter that most closely bring the spectra together is $\omega/\langle \omega_c \rangle$. This is indeed observed in \autoref{fig:qw-scaling}. In the left panel we have plotted the resulting quenching factor for the different media given a fixed \emph{initial} $\omega_c = \hat q_0 L^2/2$. In the right panel, the suppression factors are plotted for the same \emph{average} $\langle \omega_c \rangle$. Note that the Bjorken model additionally depends on the initial time $t_0$, or more precisely the ratio $\tau_0 /\tau = t_0/L$. This gives further support for the numerical scaling laws first discussed in \cite{Salgado:2002cd,Salgado:2003gb}.

\subsection{Jet suppression from soft scaling}
\label{sec:suppression-scaling}
 Let us now return the results found in Secs.~\ref{subsec:expo} and \ref{subsec:bjorken} regarding the scaling features of the rates for the different medium profiles to gain further analytical insight into the results for the medium-induced gluon distribution and suppression factor $\Raa$. We have extracted these features by only retaining the ``singular'' parts of the spectrum. Hence, the full solution that includes finite-$z$ and finite length effects should be expected to deviate. However, we expect the qualitative features to be visible in the numerical results that will be discussed in detail below.

For the exponential case, \autoref{eq:rate-expo-approx} indicates that the solution to the rate equation is given by \autoref{eq:rate-equation-solution} with $\bar \alpha \to 2 \bar \alpha$ or $\hat q_0 \to 4 \hat q_0$, namely
\beq
D(x,\tau) \approx \frac{2 \bar \alpha \tau}{\sqrt{x}} \rme^{-4\pi \bar \alpha^2 \tau^2} \,,
\eeq
where we focus on the small-$x$ regime where finite-length corrections should be smaller. This also implies that, for the same $\bar \alpha$ and $\hat q_0$, the ratio of suppression factors is,
\beq
\label{eq:raa-ratio-static-expo}
\frac{\Raa^{\rm exp}}{\Raa^{\rm static}} \simeq \exp\left[-2 \bar \alpha \sqrt{\pi \hat q_0 L^2 (n-1)/\pT} \right] \,.
\eeq
We therefore expect that all effects of the expansion can be absorbed into the proper rescaling of the parameters.

For the Bjorken scenario, the situation is slightly more complicated. However, our result for the ``singular'' rate \ref{eq:rate-Bjork-approx-2} indicates that we can write the final solution for the medium-induced gluon distribution as
\beq
D(x,\tau) 
\approx \frac{2 \bar \alpha \sqrt{\tau_0\tau}}{\sqrt{x}} \rme^{-4\pi \bar \alpha^2 \tau_0\tau} \,,
\eeq
in the small-$x$ regime. We note that the evolution time is now $\tau_\eff =2\sqrt{\tau_0 \tau}$ rather than $\tau$ itself. Interestingly, this gives a difference in suppression factors to the static case as
\beq
\label{eq:raa-ratio-static-Bjorken}
\frac{\Raa^{\rm Bjork}}{\Raa^{\rm static}} \simeq \exp\left[-2 \bar \alpha \sqrt{\pi \hat q_0 L^2 (n-1)/\pT} \left(2 \sqrt{\frac{t_0}{L}} -1 \right) \right] \,.
\eeq
Due to the additional, explicit dependence on the ratio $t_0/L$, we conclude that in the case of the Bjorken expansion there is not universal way of rescaling the parameters to arrive at the results of the static medium. Instead, approximate scaling can be achieved for given values of the medium parameters, including $t_0$.

\subsection{Numerical results on jet suppression}
\label{sec:suppresion-numerics}

As a final step, we proceed to calculating the jet suppression factor based on the medium-induced gluon distribution function obtained in \autoref{sec:rate-eq}. 
To obtain the results from numerical calculations, 
we include only one parton species (gluons) in the hard spectrum with $n=5.6$ \cite{Spousta:2015fca} and a fixed $\alpha_s = 0.14$. The distribution $D(x,\tau)$ is found by a numerical solution to \autoref{eq:RateEquation-generic}, as described in \autoref{sec:rate-eq}. 
 Due to above described limitations, the $\Raa$ calculated here is just a proxy for the nuclear modification factor, $R_\mathrm{AA}$, measured by experiments. To make 
this explicit we use symbol $\Raa$ instead of $R_\mathrm{AA}$ in this paper. 
 Finally, we mention that, in our current numerical implementation of the Bjorken model, we have not implemented the exact dependence on the rescaling of initial 
$\tau_0$, i.e. $D(x,\sqrt{x}\tau,\sqrt{x}\tau_0)\approx D(x,\sqrt{x}\tau,\tau_0)$.
 The uncertainty of the resulting distributions is calculated by varying the $\alpha_s$ parameter by 10\%. 


\begin{figure}
\centering
\includegraphics[width=0.5\textwidth]{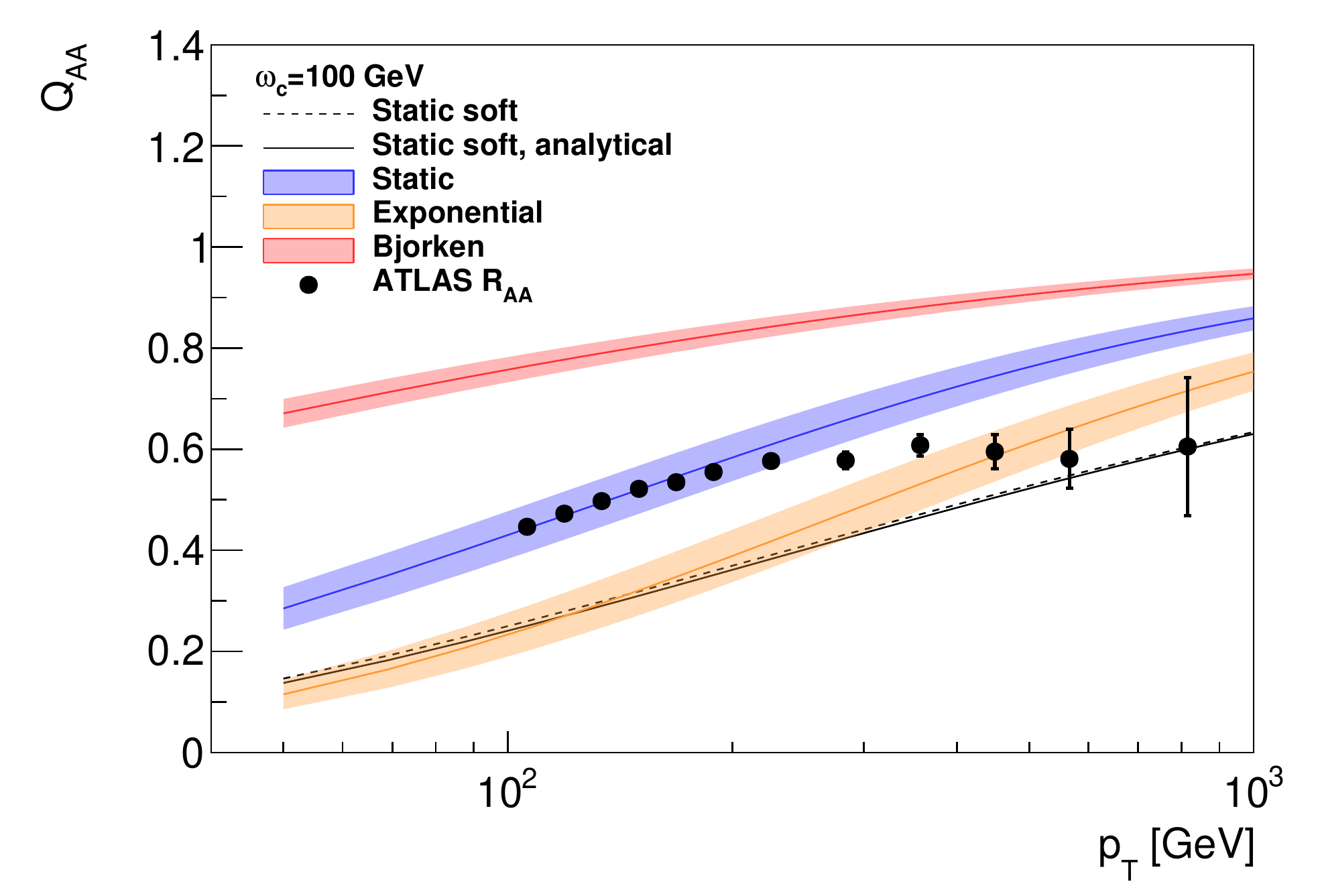}%
\includegraphics[width=0.5\textwidth]{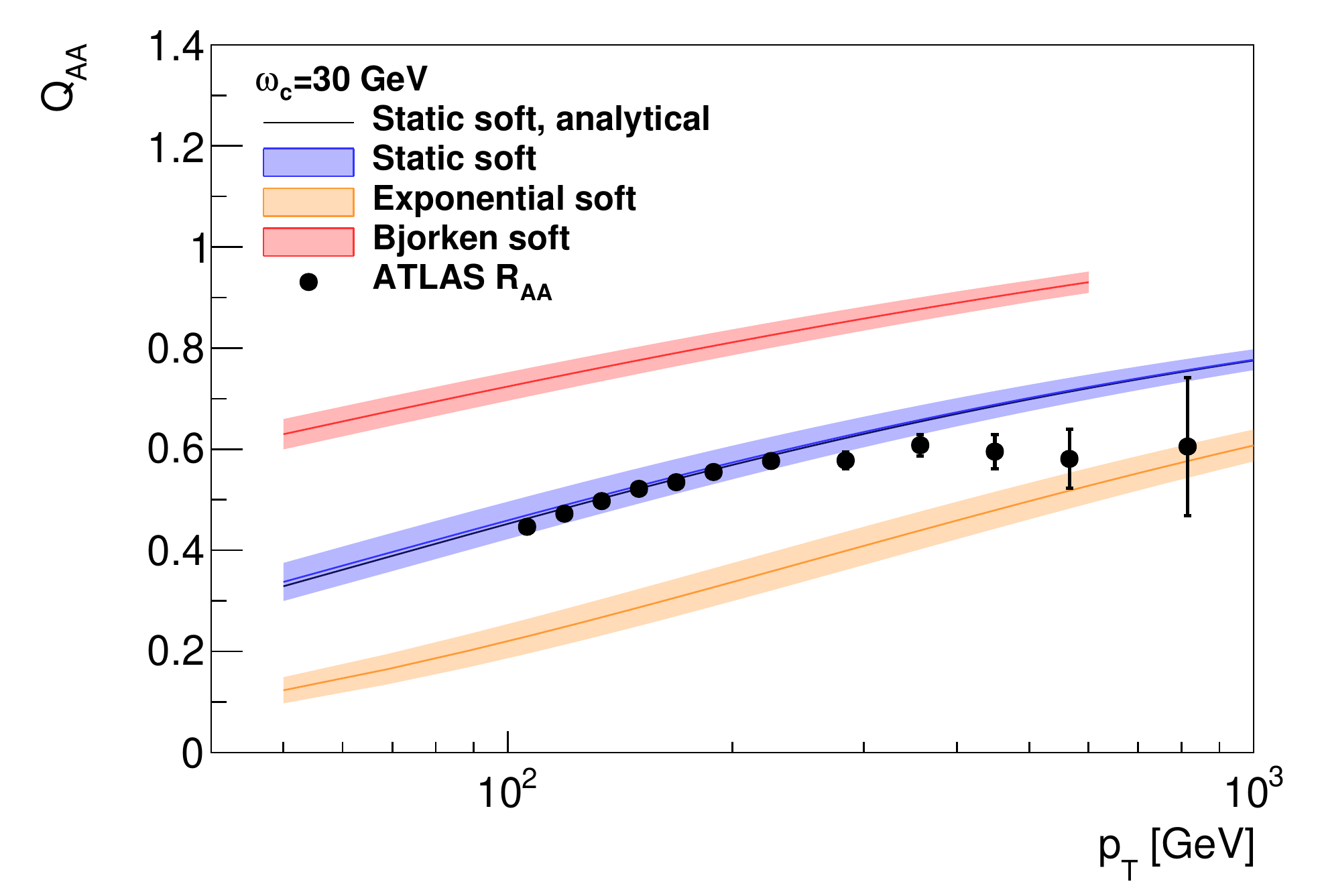}
\caption{
  The jet suppression factor $\Raa$ for four types of medium expansion calculated numerically: static soft (dashed line), static (blue), 
exponential (orange), and Bjorken (red), and the soft limit of the static medium calculated analytically (full line). 
  The $\Raa$ calculated using full kernels for $\omega_c=100$~GeV {\it (left)}.
  The $\Raa$ calculated using singular kernels for $\omega_c=30$~GeV {\it (right)}.
  Uncertainty bands correspond to 10\% variation of the value of $\alpha_s$. 
  ATLAS data taken from \cite{Aaboud:2018twu}.
 }
\label{fig:raa1}
\end{figure}

  We start our discussion by fixing a common reference value of the medium parameters. As seen before, the amount of jet suppression depends only on the energy scale 
$\omega_c\equiv\hat q_0 L^2/2$, see \autoref{sec:suppresion-qw}, up to a factor of $\bar \alpha^2$. Hence, we fix two values of $\omega_c$ for the gluon distribution 
evolved in a static medium with the full and singular kernels that lead to realistic values of the jet suppression factor for the two cases, respectively. For the 
distribution evolved with the full kernel this corresponds to $\omega_c = 100$ GeV and for the distribution evolved with the singular kernel we find $\omega_c = 30$ 
GeV.
  We show the resulting $\Raa$ distributions in \autoref{fig:raa1}, where in the left panel we plot the jet suppression factor for different media evolved using full 
kernels and, in the right panel, evolved with singular kernels.
Using a reference value of $\hat q_0 = 0.2$ GeV$^3$, from this we can extract the path-length in the medium to be $L=6.3$~fm and $3.5$~fm in these two cases.
A large difference can be seen for different media due to the varying rate of expansion in \autoref{fig:raa1}. In order to guide the eye, we have also plotted experimental data for high-$p_T$ (anti-$k_t$, $R=0.4$) jet suppression  \cite{Aaboud:2018twu}. 
Note that, for the Bjorken medium, we also have to fix the ratio $\tau_0/\tau = \sqrt{t_0/L}$ which is chosen to be $\tau_0/\tau=0.03$.

\begin{figure}
\centering
\includegraphics[width=0.32\textwidth]{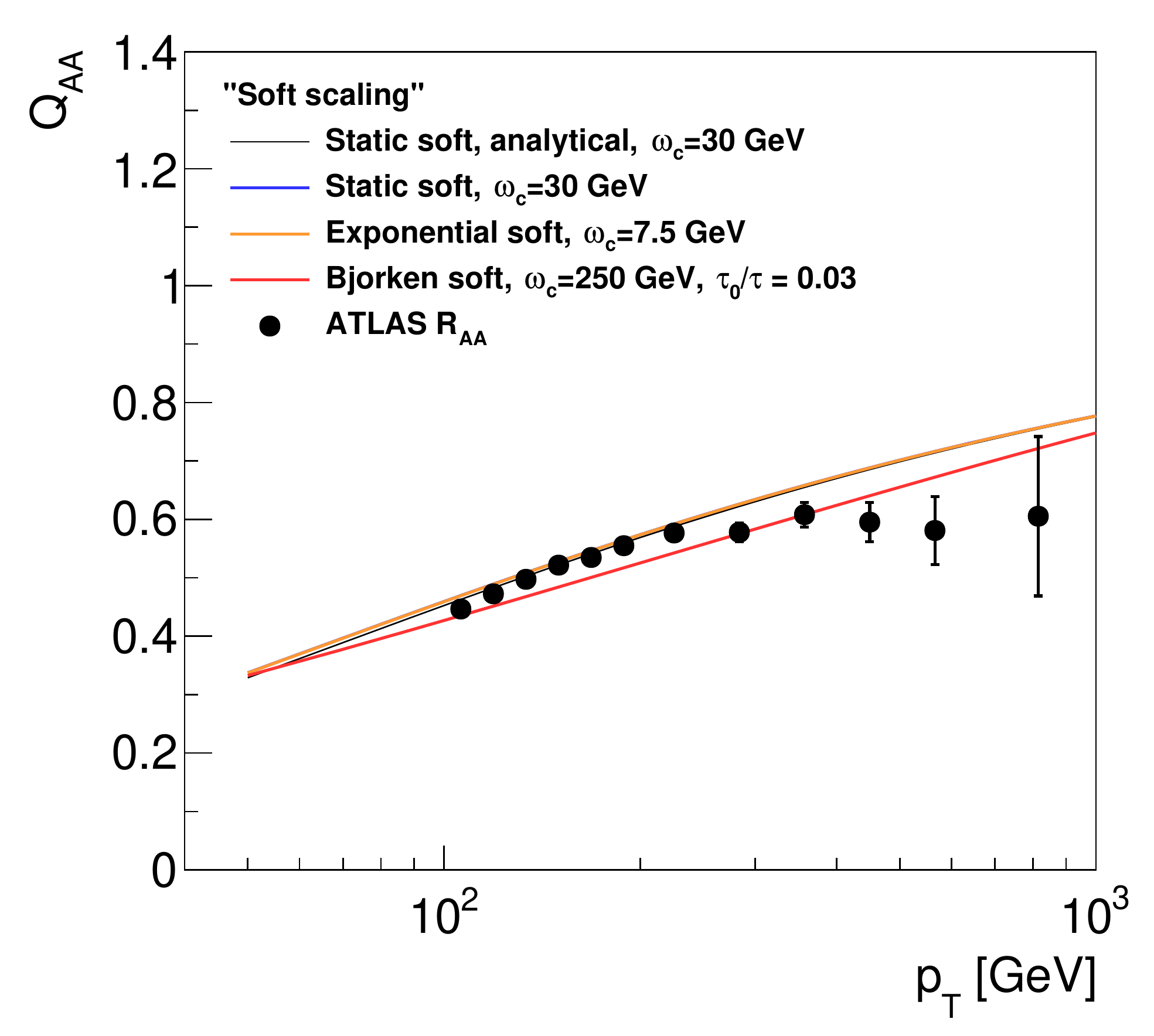}
\includegraphics[width=0.32\textwidth]{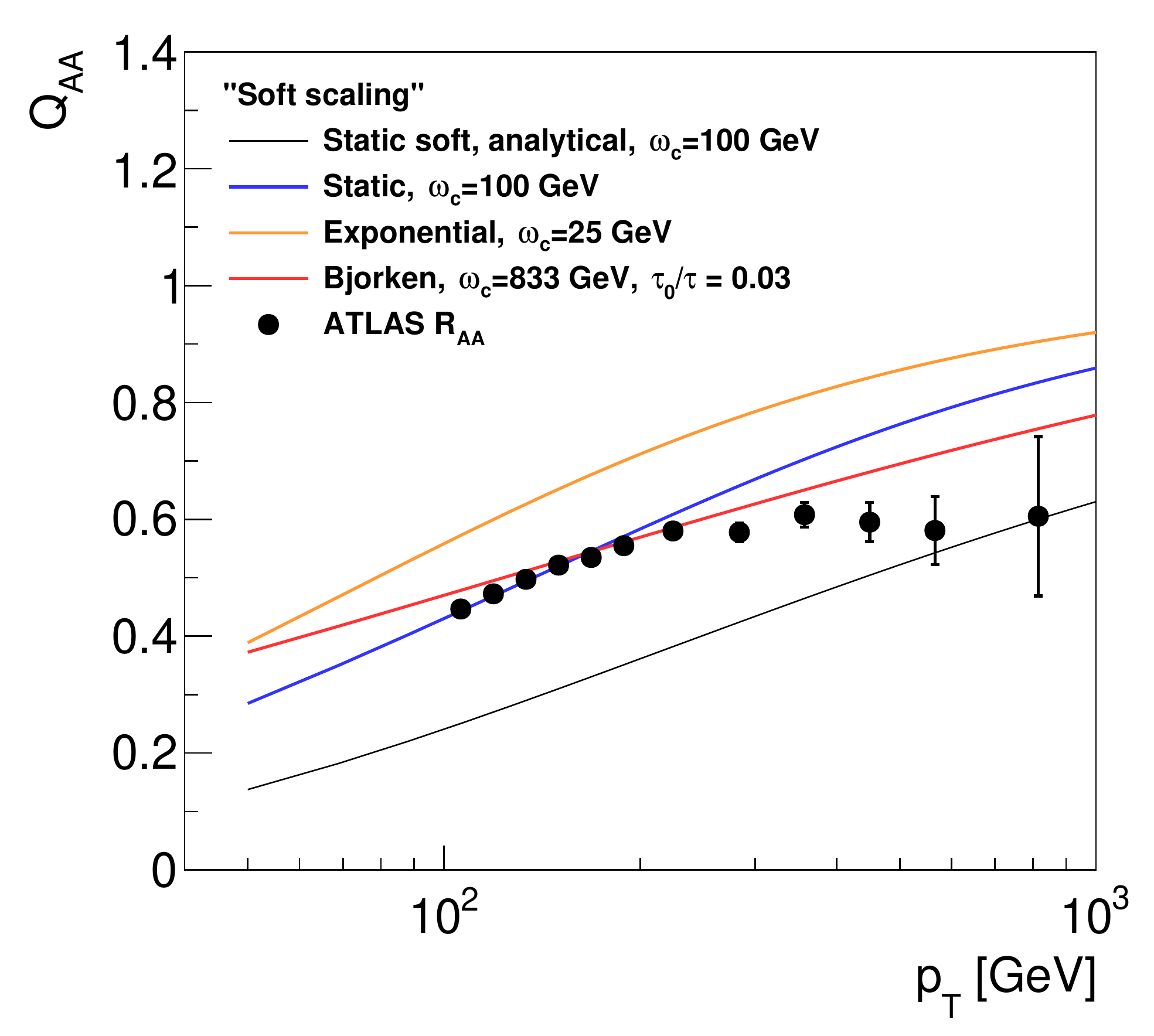}
\includegraphics[width=0.32\textwidth]{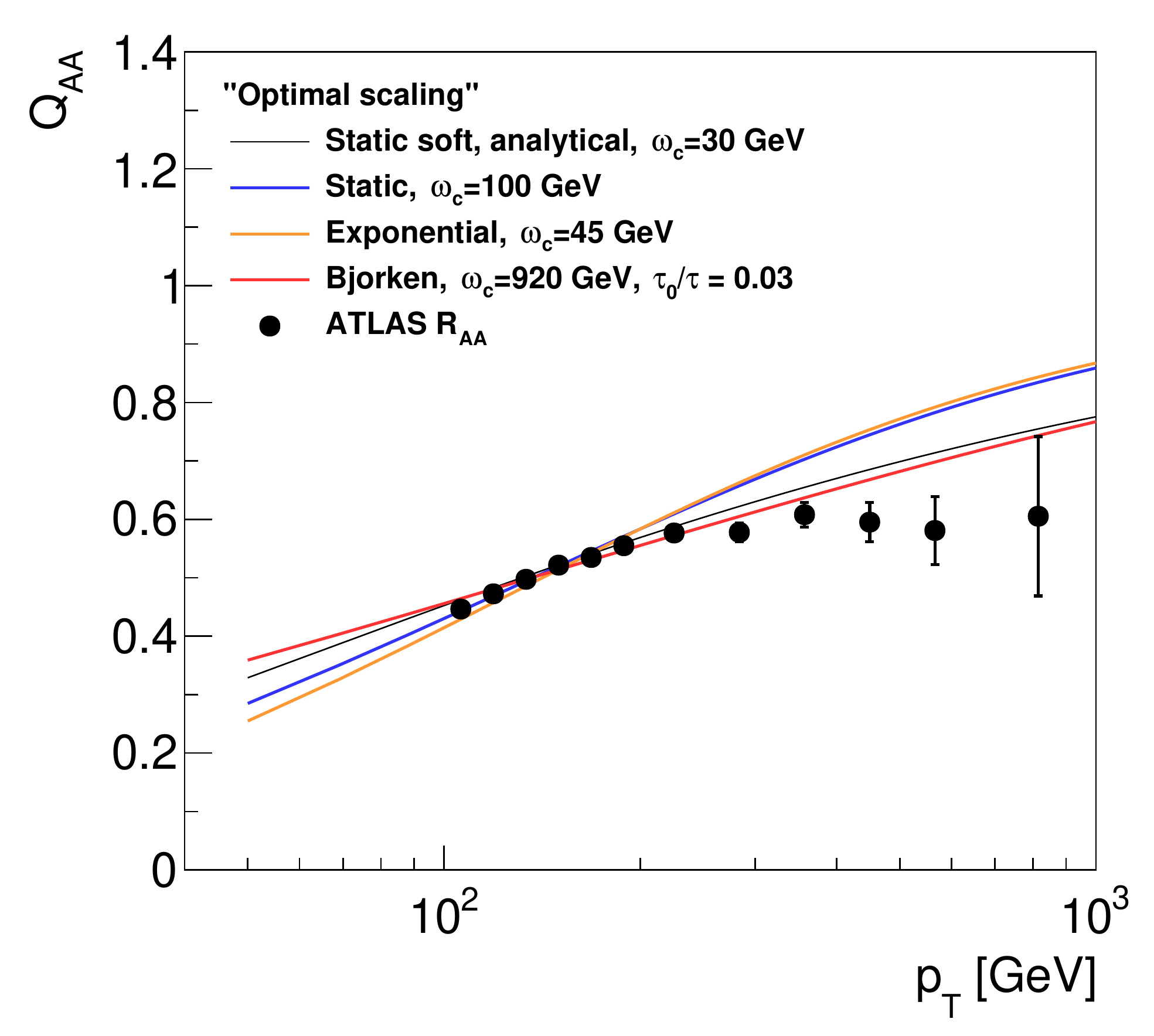}
\caption{
   The jet suppression factor $\Raa$ for three types of medium expansion calculated numerically, static (blue), exponential (orange), and Bjorken
(red), and the soft limit of the static medium calculated analytically (full line). ATLAS data taken from \cite{Aaboud:2018twu} (full closed markers).
  The $\Raa$ calculated using singular splitting kernels, the choice of $\omega_c$ values is given by scaling of $D(x)$ distributions in the soft limit and is defined by \autoref{eq:omega-eff} {\it (left)}.
  The $\Raa$ calculated using full splitting kernels, the choice of $\omega_c$ values is given by scaling of $D(x)$ distributions in the soft limit and is defined by \autoref{eq:omega-eff} {\it (middle)}.
  The $\Raa$ calculated using full splitting kernels, the choice of $\omega_c$ values is given by $\chi^2$ minimization of deviation between the caculated $\Raa$ and the data {\it (right)}. 
  }
\label{fig:raa2}
\end{figure}


  Now, we can evaluate the scaling properties of the $\Raa$. To explore the impact of the soft scaling discussed in \autoref{sec:suppression-scaling} for the analytical 
calculation of the $\Raa$ we fix the $\omega_c$ value for the static case to be the same as discussed in the previous paragraph but we replace $\omega_c$ values for the 
exponential case and Bjorken case by $\omega_\text{eff}$ values defined in \autoref{eq:omega-eff}.
  First we perform the calculation using the singular rates (\autoref{eq:rate-static-soft} for the static case, \autoref{eq:rate-expo-approx} for the exponential case 
and \autoref{eq:rate-Bjork-approx} for the Bjorken case). Results are plotted in the left panel of \autoref{fig:raa2} ($\omega_c$ values are given in the plot). We can 
see that the scaling of rates leads to a scaling of $\Raa$ in the soft limit. For a given choice of $\tau_0/\tau$, this scaling approximately holds also for the Bjorken 
case where there is no universal way of rescaling the parameters as discussed in \autoref{sec:suppression-scaling}.
  The calculation using singular rates can be compared with the calculation using the full rates, plotted in the middle panel of \autoref{fig:raa2} ($\omega_c$ values 
are shown on the plot, their ratio with respect to the static case is the same as in the left panel). Here the scaling is clearly broken leaving us with substantial 
difference between the magnitude of the $\Raa$ for the exponential case and static case. We can see that some deviation from the scaling is present for the Bjorken case 
as well, although it is less substantial than for the exponential case.

Finally, we evaluate the optimal value of $\omega_c$ for which we obtain the minimal differences among $\Raa$ distributions. This is done by $\chi^2$ minimization 
procedure which minimizes the deviation between the numerical calculations and the measured data from \cite{Aaboud:2018twu}. The result is shown in the right panel of 
\autoref{fig:raa2} along with the $\omega_c$ values. Good, but not perfect scaling of $\Raa$ is achieved by the minimization. We can see that the values of $\omega_c$ 
naturally deviate from those reported in the middle panel of \autoref{fig:raa2}. We can see that for the exponential scale the optimal scaling is very close to the 
factor of two discovered in the ``quenching weight limit'' as discussed in \autoref{sec:suppresion-qw}. On the contrary, the value of $\omega_c$ for the Bjorken scaling 
is almost a factor of two smaller compared to the value obtained in \autoref{sec:suppresion-qw} when scaling by the average $\langle \omega_c \rangle$. While this 
supports the previously discovered scaling by the average $\langle \omega_c \rangle$ for the exponential case, it shows that the rapidly decaying Bjorken type of the 
medium is very sensitive to the choice of parameters and no scaling works properly in this case.

To summarize the scaling properties for different types of medium expansion we provide in \autoref{tab:q} values of $\hat q_0$ calculated from $\omega_c$ values for the 
reference value of $L=6.3$~fm and for different scaling discussed in this section.

\begin{table}[]
\centering
\begin{tabular}{|c|c|c|c|}
\hline
$\hat q_{0}$ [GeV$^3$] & static & exponential & Bjorken \\
\hline
\hline
no scaling       & 0.2 & 0.2  & 0.2  \\ \hline
soft scaling     & 0.2 & 0.05 & 1.66     \\ \hline
optimal scaling  & 0.2 & 0.09 & 1.84     \\ \hline
scaling by $\langle \omega_c \rangle$ & 0.2 & 0.1 & 3.33 \\ \hline
\end{tabular}
\caption{Table showing a comparison of the values of the jet quenching coefficients at initial time $t_0$, $\hat q_0$, for the different medium profiles and for different types of scaling discussed in \autoref{sec:moments}. 
The first row corresponds to the reference values used in the left panel of \autoref{fig:raa1}. 
The second row corresponds to the rescaled values using the soft scaling defined by \autoref{eq:omega-eff} (middle panel of \autoref{fig:raa2}). 
The third row corresponds to the rescaled values obtained from the optimal scaling defined by the $\chi^2$ minimization of the differences between the theory and the data (right panel of \autoref{fig:raa2}).
The last row corresponds to the rescaled values obtained from the scaling by $\langle \omega_c \rangle$ discussed in \autoref{sec:suppresion-qw} (right panel of \autoref{fig:qw-scaling}). 
The values are calculated using the reference value of $L=6.3$~fm.
}
\label{tab:q}
\end{table}
\section{Conclusions \& outlook}
\label{sec:conclusions}

  Three types of the medium expansion were studied within the framework of multiple soft scattering, namely the static medium, exponentially decaying medium and 
Bjorken-like expanding medium. The spectra and rates of induced gluon emissions were evaluated and the distribution of medium-induced gluons were calculated using the 
evaluation of in-medium evolution (\autoref{eq:RateEquation-generic}) with splitting kernels obtained from rates of induced gluon emissions. Single-inclusive gluon 
distributions were then used to calculate the jet suppression factor $\Raa$, see \autoref{eq:suppression-factor-1}.

A universal behavior of splitting kernels is derived for different medium expansions in the soft gluon regime (see \autoref{eq:rate-static-soft}, \autoref{eq:rate-expo-approx}, and \autoref{eq:rate-Bjork-approx}). Using these kernels, It is shown that the impact of medium expansion on the resulting jet $\Raa$ in this soft limit can be absorbed into the proper rescaling of the parameters for the exponential case (see \autoref{eq:raa-ratio-static-expo}). For the Bjorken expansion, the onset of scaling is additionally sensitive to the initial time $t_0$. Therefore the resulting jet suppression will have a sensitivity to the ratio $t_0/L$ (see \autoref{eq:raa-ratio-static-Bjorken}).

For the full evolution in time and full phase-space of the radiation, the results are obtained by a numerical solution of the evolution equation. The evolved 
distributions are shown to obey a $1/\sqrt{x}$ scaling for all the studied types of expansion signaling a universal behavior with reduced sensitivity to the details of 
the medium expansion. However, it was found that the impact of the medium expansion on $\Raa$ cannot be scaled out in the same way as in the soft limit 
(see \autoref{fig:raa2} middle). A better agreement was found by using a phenomenologically motivated value of averaged over the path length of the jet $\langle \hat q\rangle$. However, for the Bjorken expansion we are still sensitive to the onset of quenching through the ratio $t_0/L$, which spoils the universal scaling features.
Hence, the details of the high-$x$ behavior of $D(x,\tau)$, that ultimately drive the quenching factor $\Raa$, are still sensitive to 
the details of medium expansion.

Values of quenching parameter at initial time $\hat q_{0}$ (or alternatively values of $\omega_c$ for a fixed medium length) that minimize the differences in the jet 
$\Raa$ among the different types of the expansion were also found (see \autoref{fig:raa2} right and \autoref{tab:q}). For the exponential expansion these optimal values 
are very close to the average values suggesting validity of scaling discussed in \cite{Salgado:2002cd,Salgado:2003gb}. On the contrary the rapidly decaying Bjorken type 
of the medium is very sensitive to the choice of parameters and no scaling works properly in this case. These results clearly indicate the importance of taking into 
account the medium expansion in precise modeling of the jet quenching phenomenon. Furthermore, our results shed light on the relation between jet suppression and the amount of energy deposited in the medium which are related to the high-$x$ and low-$x$ parts of the distribution, respectively.  These aspects could prove very important for more involved observables, such as the  suppression factor of reconstructed jets and jet substructure observables, and we plan to study them further in the context of dynamically evolving media.

The extracted values of the initial $\hat q_0$ obtained in this paper cannot be taken at face value, given the fitting procedure described above. Several improvements, 
such as including proper quark and gluon jet fractions \cite{Spousta:2015fca}, using a comprehensive emission rate \cite{Mehtar-Tani:2019ygg,Feal:2019xfl}, accounting 
for quark and gluon coupled induced branching \cite{Mehtar-Tani:2018zba} and including the effect of in-medium jet fragmentation (Sudakov suppression) 
\cite{Mehtar-Tani:2017web}, are planned to be included for future phenomenological applications of other observables, such as $v_2$ at high-$p_T$.

\section*{Acknowledgments}
 We would like to thank the referee of our paper, whose comments and questions helped improve the presentation of our results.
  KT is supported by a Starting Grant from Trond Mohn Foundation (BFS2018REK01) and the University of Bergen. 
  CAS is supported by Ministerio de Ciencia e Innovaci\'on of Spain under project FPA2017-83814-P; Unidad de Excelencia Mar\'ia de Maetzu under project MDM-2016-0692;
ERC-2018-ADG-835105 YoctoLHC; and Xunta de Galicia (Conseller\'ia de Educaci\'on) and FEDER.
  SPA and MS are supported by Grant Agency of the Czech Republic under grant 18-12859Y, by the Ministry of Education, Youth and Sports of the Czech Republic under grant 
LTT~17018, and by Charles University grant UNCE/SCI/013.

\bibliographystyle{jhep}
\bibliography{draft-18}

 \end{document}